\pdfoutput=1
\documentclass[journal]{IEEEtran}

\IEEEoverridecommandlockouts 

\usepackage{amsmath,amssymb,amsfonts,amsthm,commath,esint}
\usepackage[numbers,sort&compress]{natbib}

\usepackage{mathtools}
\usepackage[nolist,nohyperlinks]{acronym}
\usepackage{color}
\usepackage{multirow}
\usepackage{enumitem}
\usepackage{physics}
\usepackage{optidef}
\usepackage{mleftright}
\usepackage{cuted}
\usepackage{algorithm2e}
\usepackage{microtype}
\usepackage{bm}
\usepackage{siunitx}
\usepackage{adjustbox}

\SetKwInput{kwInit}{Initialization}

\ifCLASSOPTIONcompsoc
    \usepackage[caption=false, font=normalsize, labelfont=sf, textfont=sf]{subfig}
\else
\usepackage[caption=false, font=footnotesize]{subfig}
\fi

\makeatletter
\def\blfootnote{\xdef\@thefnmark{}\@footnotetext}
\makeatother

\begin{acronym}
\acro{BS}{base station}
\acro{CMS}{canonical minimum scattering}
\acro{DMA}{dynamic metasurface antenna}
\acrodefplural{DMA}[DMAs]{dynamic metasurface antennas}
\acro{LIS}{large intelligent surface}
\acro{LoS}{line-of-sight}
\acro{mMIMO}{massive multiple-input multiple-output}
\acro{PEC}{perfect electric conductor}
\acro{RF}{radio-frequency}
\end{acronym}

\newcommand{\es}[1]{\sin{\left(#1\right)}}
\newcommand{\ec}[1]{\cos{\left(#1\right)}}
\newcommand{\p}[1]{\left(#1\right)}

\renewcommand{\vec}[1]{\mathbf{\lowercase{#1}}}
\newcommand{\nvec}[1]{\hat{\mathbf{\lowercase{#1}}}}
\newcommand{\mat}[1]{\mathbf{\uppercase{#1}}}

\newcommand{\Int}{\int\limits}

\newcommand{\e}[1]{\mathrm{e}^{#1}}

\NewDocumentCommand{\evalat}{sO{\big}mm}{%
  \IfBooleanTF{#1}
  {\mleft. #3 \mright|_{#4}}
  {#3#2|_{#4}}%
}

\DeclarePairedDelimiterXPP\Aver[1]{\mathbb{E}}{[}{]}{}{

#1
}

\newcommand{\review}[1]{{#1}}

\usepackage{fancyhdr}

\begin{document}

\title{Electromagnetic Based Communication Model for Dynamic Metasurface Antennas}

\author{{Robin Jess Williams,
        Pablo Ram\'irez-Espinosa,
        Jide Yuan and Elisabeth De Carvalho}
\thanks{Manuscript received XXXXX XX, 2021; revised XXXXX XX, 2021. The review of this paper was coordinated by XXXX.}
\thanks{The authors were with Department of Electronic Systems, Connectivity Section (CNT), Aalborg University, Denmark. E-mail: \{rjw, pres, jyu, edc\}@es.aau.dk.}
\thanks{This work has been supported by the Danish Council for Independent Research under grant DFF-701700271.}
}

\maketitle
\thispagestyle{fancy}

\begin{abstract}
Dynamic metasurface antennas (DMAs) arise as a promising technology in the field of massive multiple-input multiple-output (mMIMO) systems, offering the possibility of integrating a large number of antennas in a limited ---and potentially large--- aperture while keeping the required number of radio-frequency (RF) chains under control. Although envisioned as practical realizations of mMIMO systems, DMAs represent a new paradigm in the design of signal processing techniques (such as beamforming) due to the constraints inherent to their physical implementation, for which no complete models are available yet. In this work, we propose a complete and electromagnetic-compliant narrowband communication model for a generic DMA based system. Specifically, the model accounts for: \textit{i)} the wave propagation and reflections throughout the waveguides that feed the antenna elements, \textit{ii)} the mutual coupling both through the air and the waveguides, and \textit{iii)} the insertion losses. Also, we integrate the electromagnetic model in the conventional digital communication model, providing a complete and useful framework to design and characterize the performance of these systems. Finally, the accuracy of the model is verified through full-wave simulations. 
\end{abstract}

\section{Introduction}\label{sec:introduction}
Holographic beamforming and \ac{LIS} are two concepts which have gained substantial popularity in the research community during last years. Both refer to a radically new way of designing arrays of antennas, where the classical concept of array is replaced by a continuous electromagnetic surface able to transmit and receive radio waves, pushing the theoretical benefits of \ac{mMIMO} to their limits \cite{Bjornson2019}. The theoretical performance and potential for communications of these electromagnetic sheets  have been characterized in \cite{Hu2018, Dardari2020}, although several challenges need to be addressed for realistic and realizable designs and implementations, mainly because of the large physical aperture, real-time processing and interconnection issues \cite{Tataria2020}.

In this context, \acp{DMA} have been recently proposed as a feasible implementation for \ac{LIS} \cite{Smith2017,Sleasman2016, Shlezinger2021}. Formally, a \ac{DMA} is conceived as a collection of sub-wavelength metamaterial elements deployed on top of a guiding structure (e.g., microstrip guides) \cite{Schlezinger2019}. By introducing some simple semiconductor devices into each element, their electromagnetic properties can be modified and thus the whole surface becomes reconfigurable. Stacking several of these structures leads to a large aperture antenna, and the sub-wavelength elements allow to emulate the continuous electromagnetic surface, achieving the envisioned idea of \ac{LIS}.  

To properly design and optimize communications systems based on \acp{DMA}, an accurate and realistic --- albeit tractable --- model is of capital importance. In striking contrast to conventional \ac{mMIMO} systems, the particular physical characteristics of the \ac{DMA} give rise to different phenomena which need to be considered:
\begin{enumerate}[label={\alph*)}]
    \item Once some power is introduced in the waveguide, the electromagnetic field arriving at each radiating element is affected by the propagation inside the guide itself and the field radiated by the previous elements. 
    \item To emulate a continuous surface, the metamaterial elements are intended to be closely spaced. Then, the interaction or coupling between them should be taken into account. Since the elements are placed on top of the guide, this interaction occurs through the air but also through the waveguide, both with different characteristics \cite{Pulido2017,Pulido2018}.
    \item The reconfigurability of the surface implies that the impedances of the elements vary over time. This has a non-negligible twofold impact: \textit{i)} the equivalent impedance of the waveguide changes, and so the insertion losses; and \textit{ii)} the interaction between the different radiating elements also varies.
\end{enumerate}

From a communication viewpoint, the three aforementioned effects translate into an equivalent channel that not only depends on the propagation environment but also on the \ac{DMA} design and its specific --- and momentaneous --- configuration. Moreover, for a fixed transmitter power, the actual power delivered to the \ac{DMA} is also affected by the chosen configuration. Hence, these effects should be modeled and considered when designing the optimization algorithms for, e.g., beamforming, in order to render realistic solutions. 

Looking at the literature, some works have carried out partial analyses of \acp{DMA}. A magnetic dipole model for the radiating elements is introduced in \cite{Johnson2014}, and later used in \cite{Johnson2015, Pulido2016, Pulido2017, Pulido2018, Yoo2019_Planar, Smith2017}, allowing to characterize the system based on their effective polarizability. The interaction between the metamaterial elements and the waveguide is analyzed in \cite{Pulido2017}, and the same analysis is extended to incorporate the coupling between elements in \cite{Pulido2018}. Following the same approach, the same authors focused on the effect of parasitic elements in linear \acp{DMA} in \cite{Pulido2016}, and on the analysis of planar metasurface antennas in \cite{Yoo2019_Planar}. 

A first step in the analysis of the whole \ac{DMA} system based on the dipole model is given in \cite{Smith2017}, where some basic beamforming techniques are also presented to exploit the potential of this system. However, neither the interaction between the elements and the waveguide, nor reflections inside the microstrip, are considered (although they are analyzed separately in \cite{Pulido2017} and \cite{Pulido2016}, respectively). A more thorough beamforming design is presented in \cite{Schlezinger2019} for the uplink and \cite{Wang2019} for the downlink, where the waveguide effect is modeled as a discrete filter (how to obtain the coefficients is not specified, though) and the coupling between the elements is ignored. In \cite{Yoo2019}, the performance of \ac{DMA} over clustered channels is analyzed, with the \ac{DMA} characterized through full-wave simulations, whilst their capabilities in the near field are simulated in \cite{Yurduseven2017}.

Nevertheless, despite the potential of \ac{DMA} in communications, a complete and tractable model is still missing. The dipole framework presented in \review{\cite{Johnson2014, Johnson2015, Pulido2016, Pulido2017, Pulido2018, Smith2017, Yoo2019_Planar}} is a promising approach, but it is difficult to incorporate in the conventional \ac{mMIMO} communication model, and hence difficult to use for beamforming and system design. Similarly, the interaction between the waveguide, the coupling and the polarizability extraction are covered in \cite{Pulido2016}, \cite{Smith2017,Pulido2018} and \cite{Pulido2017} from an antenna theory perspective, being again hard to integrate these effects in standard communication analyses and designs. In turn, the filtering model used in \cite{Schlezinger2019, Wang2019} does not consider important effects such as coupling and interaction between elements and waveguide, so the predicted results may not be realistic.  

Aiming to fill this gap, we introduce in this work a complete circuital communication model for generic \ac{DMA} systems, which can be easily introduced as part of the \ac{mMIMO} model. Our proposal accounts for the three different characteristics of \ac{DMA} systems previously mentioned, and it arises from the electromagnetic theory and the dipole model in \cite{Pulido2017, Pulido2018, Smith2017}, being therefore coherent with the underlying physics of the system. In other words, the model is designed such that it can be used for communication performance analysis and design purposes but abstracting --- albeit including --- all the electromagnetic effects. Specifically, the contributions are: 
\begin{itemize}
    \item A complete communication model for \ac{DMA} based systems is proposed, which is written in terms of mutual admittance matrices, allowing to capture mutual coupling, propagation and reflections inside the waveguides, and insertion losses. Also, closed-form expressions are given for all the admittance matrices. 
    \item The circuital model is integrated with conventional approaches to characterize the wireless propagation environment, giving rise to an equivalent channel that can be used as a starting point for, e.g., beamforming design. Thanks to the mutual admittance formulation, any propagation model (including empirical samples) can be incorporated.
    \item The validity of the model is checked through full-wave simulations, showing an accurate matching between theoretical and simulated results. 
\end{itemize}

 In summary, this paper establishes a bridge between the works on \acp{DMA} from the antenna theory viewpoint and the communication and signal processing perspective, being envisioned as a tool to pave the way for more complete and realistic analyses. While parts of the proposed model are applicable for wideband analysis, this work assumes a narrowband channel and a model for the metamaterial elements that only captures narrowband scattering characteristics. The proposed model is therefore primarily applicable for narrowband analysis.

The remainder of this paper is given as follows. The system under consideration and the circuital model are introduced in Section \ref{sec:systemModel}. Section \ref{sec:InsertionLosses} derives the basic input-output relationships of the model and characterizes the insertion losses. The analysis of the waveguides and the characterization of the \ac{DMA} elements is carried out in Section \ref{sec:DMAanalysis}, whilst the coupling between the users and the propagation channel matrix are detailed in Section \ref{sec:UserCoupling}. The mutual admittance model is merged with the conventional digital \ac{mMIMO} communication model in Section \ref{sec:MIMOmodel}, and the validation with full-wave simulations is presented in Section \ref{sec:Simulations}. Finally, some conclusions are drawn in Section \ref{sec:Conclusions}. 

\textit{Notation:} Vectors and matrices are represented by bold lowercase and uppercase symbols, respectively.  $\nabla\times(\cdot)$ is the curl operator, $\nabla(\cdot)$ is the gradient operator, $(\cdot)^T$ denotes the matrix transpose, $(\cdot)^H$ is the transpose conjugate and $(\mat{A})_{j,k}$ is the $j,k$-th element of $\mat{A}$. Also, $\vec{x}\times \vec{y}$ is the cross product between $\vec{x}$ and $\vec{y}$, $\mat{I}_n$ is the identity matrix of size $n\times n$; and $\hat{\vec{x}}= \begin{bmatrix}1&0&0\end{bmatrix}^T$, $\hat{\vec{y}}= \begin{bmatrix}0&1&0\end{bmatrix}^T$ and $\hat{\vec{z}}= \begin{bmatrix}0&0&1\end{bmatrix}^T$ denote the unitary vectors along the $x$, $y$ and $z$-axis, respectively. Finally, $i = \sqrt{-1}$ is the imaginary number, $\|\cdot\|_2$ is the $\ell_2$ norm of a vector, $(\cdot)^*$ indicates complex conjugate, $\mathbb{E}[\cdot]$ is the mathematical expectation and $\Re\{\cdot\}$ and $\Im\{\cdot\}$ denotes real and imaginary parts.

\section{System Model}\label{sec:systemModel}
\begin{figure}[t]
    \centering
    \includegraphics[width = \columnwidth]{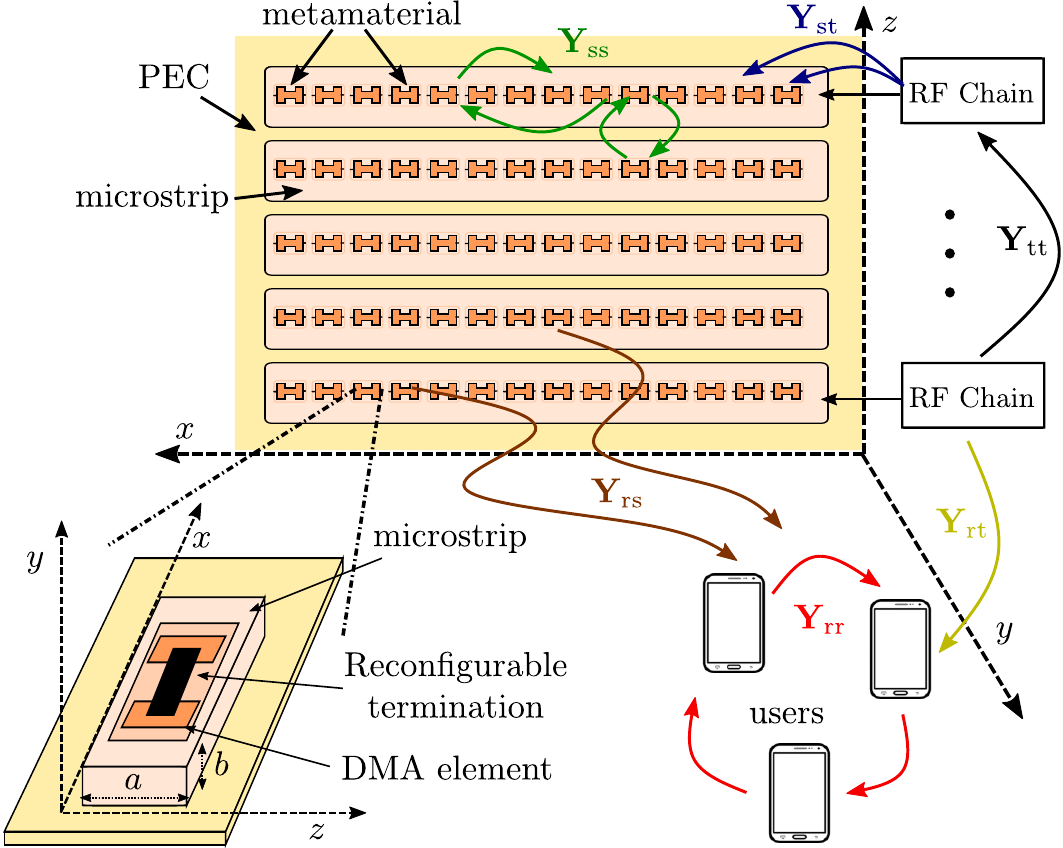}
    \caption{Graphical representation of the system model and the DMA structure. The interactions (mutual admittances) modeled by each admittance submatrix in \eqref{eq:CircuitalModel} are also depicted.}
    \label{fig:system}
\end{figure}

We consider a multi-user \ac{mMIMO} setup, in which a \ac{BS}, equipped with a \ac{DMA}, is intended to communicate with $M$ single antenna users simultaneously, i.e., we focus on the downlink, although the resulting model is also valid for the uplink. The \ac{DMA} is placed at the positive $xz$-plane of a cartesian coordinate system, whilst the users are arbitrary placed in front of it (positive values of the $y$-axis). The \ac{DMA} is composed by $N$ one dimensional waveguides of length $S_\mu$ with $L_\mu$ small antenna elements embedded in each of them, being the total number of antennas denoted by $L = L_\mu N$. The waveguides are connected to $N$ dedicated \ac{RF} chains (digital baseband precoders), while a semiconductor device is attached to each antenna element, allowing to deliberately tune and thus change the admittance of the elements \cite{Smith2017, Schlezinger2019}, as shown in Fig. \ref{fig:system}. The basic behaviour of the system is as follows: the \ac{RF} chains introduce a radio signal into the waveguides, and it propagates through them. When the signal arrives at a radiating element, part of the field inside the waveguide is scattered both inside and outside of the waveguide. The amplitude and phase of the scattered field is controlled by the termination admittance $Y_{sl}$ of the radiating element. The rest of the field remaining in the waveguide keeps propagating through, until arriving the next element where the procedure repeats.

\begin{figure}[t]
    \centering
    \includegraphics[width = \columnwidth]{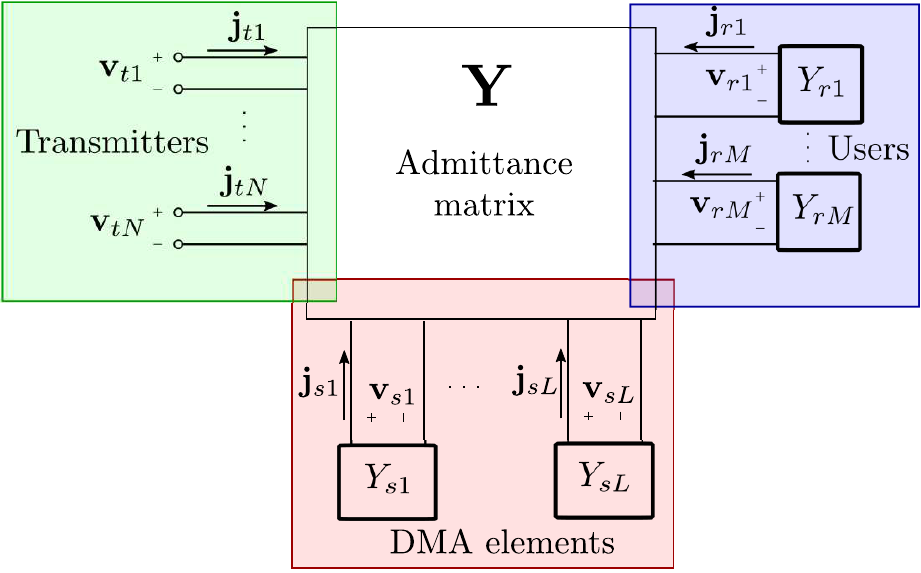}
    \caption{Multi-port model for DMA based system.}
    \label{fig:network}
\end{figure}

The described system is represented as a multi-port network in Fig. \ref{fig:network}. The ports in the left part of the figure represent the input to the system, corresponding to the $N$ different \ac{RF} chains that feed the \ac{DMA}. Each reconfigurable radiating element is then represented by another port, and the same is done for the different users. The \ac{DMA} elements are terminated in tunable admittances $Y_{sl}$ with $l=1\dots,L$, representing the reconfigurability of the surface. Also, all the users' ports are terminated in load admittances $Y_{rm}$ with $m=1\dots,M$. Since the interaction between the currents in the transmitter and the different ports only depends on the admittance matrix $\mat{Y}$, we can incorporate in this matrix effects such as mutual coupling, interaction between the \ac{DMA} elements and the waveguide, scattering between receivers and transmitter in near field, or different wireless channel conditions. Under the reciprocity assumption, i.e., the coupling between port $s$ to $q$ is the same than that from $q$ to $s$\footnote{Note that reciprocity is inherent to most passive systems, and only a very mild assumption for wireless channels.}, the system is completely described by 
\begin{align}
    \begin{bmatrix}\vec{v}_\text{t} \\ \vec{v}_\text{s} \\ \vec{v}_\text{r} \end{bmatrix} = \underbrace{\begin{bmatrix}\mat{Y}_\text{tt} &  \mat{Y}_\text{st}^T & \mat{Y}_\text{rt}^T \\ \mat{Y}_\text{st} & \mat{Y}_\text{ss} & \mat{Y}_\text{rs}^T \\ \mat{Y}_\text{rt} & \mat{Y}_\text{rs} & \mat{Y}_\text{rr} \end{bmatrix}}_{\mat{Y}}  \begin{bmatrix}\vec{j}_\text{t} \\ \vec{j}_\text{s} \\ \vec{j}_\text{r} \end{bmatrix}.  \label{eq:CircuitalModel}
\end{align}

In \eqref{eq:CircuitalModel}, $\vec{j}_\text{t}\in\mathbb{C}^{N\times 1}$, $\vec{j}_\text{s}\in\mathbb{C}^{L\times 1}$ and $\vec{j}_\text{r}\in\mathbb{C}^{M\times 1}$ represent, respectively, the magnetic currents entering the network (hereinafter named as transmitter currents), the \ac{DMA} radiating elements and the users, and are measured in volts (see Fig. \ref{fig:network}). Similarly, $\vec{v}_\text{t}\in\mathbb{C}^{N\times 1}$, $\vec{v}_\text{s}\in\mathbb{C}^{L\times 1}$ and $\vec{v}_\text{r}\in\mathbb{C}^{M\times 1}$ are the vectors containing the magnetic voltages across the corresponding ports measured in amperes. The relationship between voltages and currents is hence determined by the admittance matrix $\mat{Y}\in\mathbb{C}^{\p{N+L+M}\times \p{N+L+M}}$ measured in siemens, which can be split into different submatrices. Following standard circuital theory \cite{Pozar2012}, the admittance matrix is defined such that its elements are the mutual admittance between two ports, i.e., $(\mat{Y})_{s,q} = {v_{s}}/{j_{q}}$, with $v_{s}$ the magnetic voltage induced at port $s$ when a magnetic current $j_{q}$ is introduced in port $q$ and the rest of the ports in the network are shorted (equivalently terminating in an infinite admittance).

As shown in Fig. \ref{fig:system}, each of these submatrices capture the interaction between different ports (agents) in the system. Thus, $\mat{Y}_\text{tt}\in\mathbb{C}^{N\times N}$ represents the coupling between the different transmitter ports (ideally a diagonal matrix if the \ac{RF} chains are properly isolated), $\mat{Y}_\text{ss}\in\mathbb{C}^{L\times L}$ captures the interaction between the \ac{DMA} elements and $\mat{Y}_\text{rr}\in\mathbb{C}^{M\times M}$ does the same for the users. On the other hand, $\mat{Y}_\text{st}\in\mathbb{C}^{L\times N}$ represents the coupling between each transmitter and the reconfigurable elements, i.e., how much signal introduced by the \ac{RF} chain reaches each radiating element,  whilst  $\mat{Y}_\text{rt}\in\mathbb{C}^{M\times N}$ is the coupling between \ac{RF} chains and users. To make the following analysis tractable, we restrict our focus to \ac{DMA} elements acting as \ac{CMS} antennas \cite{Wasylkiwskyj1970} with respect to admittance parameters. However, as noted in \cite{rogers_application_1986}, simple small antennas can be made to act as \ac{CMS} antennas with proper choice of reference impedance. As a result, the admittance $\mat{Y}_\text{rt} = \mat{0}$, since it represents the coupling when all the \ac{DMA} elements are turned off (i.e., shorted)\footnote{Note that for \ac{CMS} antennas, the \ac{DMA} elements act as if they are non-existent when they're shorted, preventing therefore any radiation leakage out of the waveguides.}. Finally, $\mat{Y}_\text{rs}\in\mathbb{C}^{M\times L}$ represents the wireless channel between the users and each radiating element, and hence the effects of the propagation environment are captured here. 

In realistic systems, the only parameters that can be adjusted are the currents $\vec{j}_\text{t}$ introduced to the waveguides (equivalently, the voltages), and the terminating admittances $Y_{\text{s}l}$ of the radiating elements (i.e., the configuration of the \ac{DMA}). Hence, we are interested in characterizing the performance of the system from these inputs. The rest of parameters are imposed by the physical implementation of the \ac{DMA} and the propagation scenario.

As a final remark, note that the circuital model in \eqref{eq:CircuitalModel} is formulated in terms of admittances and magnetic currents instead of the more conventional impedance and electric currents based model \cite{Gradoni2021, Williams2020}. The rationale behind this is simplifying the analytical derivations in the following sections; since we inherit the magnetic dipole model for the \ac{DMA} elements from \cite{Pulido2017, Pulido2018, Smith2017}, it is easier to work with magnetic currents rather than electric ones without compromising the validity of the results. Henceforth, every antenna in the system is modeled as a magnetic dipole.  

\section{Input-Output Relationships}
\label{sec:InsertionLosses}
Since the inputs to the system --- i.e., the parameters that can be tuned an controlled --- are $\vec{j}_\text{t}$ and the terminating admittances $Y_{sl}$, we aim to express the output of the system, namely the induced magnetic currents $\vec{j}_\text{r}$ in the users, in terms of $\vec{j}_\text{t}$. 

From Ohm's law at both the reconfigurable elements and the users, we have 
\begin{align}
    \vec{v}_\text{s} = -\mat{Y}_\text{s} \vec{j}_\text{s}, \label{eq:ohm1}\\
    \vec{v}_\text{r} = -\mat{Y}_\text{r} \vec{j}_\text{r}, \label{eq:ohm2} 
\end{align}
where the negative sign is due to the current flowing from the negative terminal to the positive one as illustrated in Fig. \ref{fig:network}, and $\mat{Y}_\text{s}\in\mathbb{C}^{L\times L}$ and $\mat{Y}_\text{r}\in\mathbb{C}^{M\times M}$ are diagonal matrices with elements $(\mat{Y}_\text{s})_{l,l} = Y_{sl}$ and $(\mat{Y}_\text{r})_{m,m} = Y_{rm}$. Combining \eqref{eq:ohm1} and \eqref{eq:ohm2} with the first two rows of \eqref{eq:CircuitalModel}, applying the assumption of $\mat{Y}_\text{rt} = \mat{0}$, and isolating for the receiver currents yields
\begin{align}
    \vec{j}_\text{r} =& \left(\mat{Y}_\text{r}+\mat{Y}_\text{rr}-\mat{Y}_\text{rs}\left(\mat{Y}_\text{s}+\mat{Y}_\text{ss}\right)^{-1}\mat{Y}_\text{rs}^T\right)^{-1} \notag \\
    &\times\left(\mat{Y}_\text{rs}\left(\mat{Y}_\text{s}+\mat{Y}_\text{ss}\right)^{-1}\mat{Y}_\text{st}\right)\vec{j}_\text{t}, \label{eq:IO}
\end{align}
which represents the input-output relationship of the system. As expected, \eqref{eq:IO} does not only depend on the \ac{DMA} characteristics but also on the propagation environment $\mat{Y}_\text{rs}$ and the terminating admittances $\mat{Y}_\text{s}$ and $\mat{Y}_\text{r}$. 

A closer look to \eqref{eq:IO} reveals that the coupling from the users back to the \ac{BS}, i.e. the term $\mat{Y}_\text{rs}\left(\mat{Y}_\text{s}+\mat{Y}_\text{ss}\right)^{-1}\mat{Y}_\text{rs}^T$, may actually be negligible for relatively large distances between \ac{BS} and users. Since the matrix $\mat{Y}_\text{rs}$ accounts for the wireless channel and hence encapsulates the pathloss, the relative amplitude of such term decreases with twice the pathloss or, equivalently, the distance between users and \ac{BS}. With this in mind, we can apply the far field or unilateral assumption, and then \eqref{eq:IO} simplifies to
\begin{align}
    \vec{j}_\text{r} \approx \left(\mat{Y}_\text{r}+\mat{Y}_\text{rr}\right)^{-1} \left(\mat{Y}_\text{rs}\left(\mat{Y}_\text{s}+\mat{Y}_\text{ss}\right)^{-1}\mat{Y}_\text{st}\right)\vec{j}_\text{t}. \label{eq:IO_FF}
\end{align}

On the other hand, the received power at the $m$-th user is given as the time-averaged power dissipated in the receiver-attached load as \cite{Pozar2012}
\begin{align}
    P_{\text{r}m} = \frac{\Re{-(\vec{j}^H_\text{r})_m (\vec{v}_\text{r})_m}}{2} = \frac{\|(\vec{j}_\text{r})_m\|^2 \Re{\p{\mat{Y}_\text{r}}_{m,m}}}{2}. \label{eq:powerRecieved}
\end{align}

\subsection{Insertion losses}

On the transmitter side, the two metrics that are of interest in communications oriented applications are the transmitted power $P_t$ and the supplied power $P_s$, being the former the power introduced in the system by the transmitter ports in Fig. \ref{fig:network} and the latter the power that is actually delivered by the \ac{RF} chains. That is, only part of the power $P_s$ available is delivered to the network due to admittance mismatching and reflections. Specifically, the transmitted power is given by
\begin{align}
   P_\text{t} = \frac{\Re{\vec{j}_\text{t}^H \vec{v}_\text{t}}}{2} \overset{(a)}{\approx} \frac{\Re{\vec{j}_\text{t}^H \mat{Y}_\text{p}\vec{j}_\text{t}}}{2}, \label{eq:powerTransmitted}
\end{align}
where the far field approximation and the fact that $\mat{Y}_\text{rt} = \mat{0}$ are used in $(a)$. In \eqref{eq:powerTransmitted}, $\mat{Y}_\text{p} = \mat{Y}_\text{tt}- \mat{Y}_\text{st}^T\left(\mat{Y}_\text{s}+\mat{Y}_\text{ss}\right)^{-1} \mat{Y}_\text{st}$ represents the \ac{RF} chain admittance matrix when the \ac{DMA} elements are terminated by loads $\mat{y}_\text{s}$ and the users are located in far-field, i.e., $\mat{Y}_\text{p}$ is defined so that $\vec{v}_\text{t}= \mat{Y}_\text{p}\vec{j}_\text{t}$. Note that $\mat{Y}_\text{p}$ is readily obtained from the first two rows of \eqref{eq:CircuitalModel} by applying \eqref{eq:ohm1} and the unilateral assumption. 

The input admittance $Y_{\text{in},n}$ at the $n$-th \ac{RF} chain is determined by the ratio of the corresponding magnetic voltage and current as
\begin{align}
    Y_{\text{in}, n} = \frac{(\vec{v}_\text{t})_n}{(\vec{
    j}_\text{t})_n} = \frac{\sum_{m}^N (\mat{Y}_{\text{p}})_{n,m}  (\vec{
    j}_\text{t})_m}{(\vec{
    j}_\text{t})_n} \label{eq:inAdmittance}
\end{align}
and, from \eqref{eq:inAdmittance}, the reflection coefficients at the different \ac{RF} chains are given by the diagonal matrix $\mat{\Gamma}\in\mathbb{C}^{N\times N}$ with elements
\begin{align}
    (\mat{\Gamma})_{n,n} = -\frac{{Y}_{\text{in}, n}-Y_0}{Y_{\text{in}, n} + Y_0}, \label{eq:Gamma}
\end{align}
where $Y_0$ is the intrinsic admittance of the transmission line connecting each \ac{RF} chain and the corresponding \ac{DMA} waveguide (the above definition in terms of admittances is readily obtained from \cite[Eqs. (2-14)-(2-17)]{Pozar2012} by inverting the impedances). With the reflection coefficient, the relation between transmitted power $P_\text{t}$ and supplied power $P_\text{s}$ within each \ac{RF} chain is given as \cite[Eq. (2.37)]{Pozar2012}
\begin{align}
    \vec{P}_\text{s} =  \left( \mat{I}_N - \mat{\Gamma}^H\mat{\Gamma}\right)^{-1} \vec{P}_\text{t},
\end{align}
where $\vec{p}_\text{s}\in\mathbb{C}^{N\times 1}$ and $\vec{p}_\text{t}\in\mathbb{C}^{N\times 1}$ are the supplied power vector at the \ac{RF} chains and the transmitted power vector defined as $(\vec{p}_\text{t})_n = \Re{(\vec{j}^H_\text{t})_n(\vec{v}_\text{t})_n}/2$. Finally, the total power supplied to the system is straightforwardly obtained as
\begin{align}
   P_\text{s} = \sum_{n=1}^N(\vec{p}_\text{s})_n  \overset{(a)}{=} \frac{\Re{\vec{j}_\text{t}^H \left( \mat{I}_N - \mat{\Gamma}^H\mat{\Gamma}\right)^{-1}\mat{Y}_\text{p}\vec{j}_\text{t}}}{2}, \label{eq:powerSupplied}
\end{align}
in which, once again, the unilateral assumption is considered in $(a)$. 
Interestingly, note that the reflection coefficient matrix $\mat{\Gamma}$, and hence the actual power $P_s$ introduced in the system, depends on both the currents introduced in adjacent waveguides and the termination admittance of the radiating elements (and thus the \ac{DMA} configuration) through $\mat{Y}_\text{p}$. This implies that, when selecting a specific configuration to, e.g., render a beam in an arbitrary direction, we are also changing the effective power that is supplied to the system. Therefore, considering  \eqref{eq:powerSupplied} when designing \ac{DMA} based systems seems to be of relevance. 

As a final remark, note we are analyzing the system performance and design in terms of the transmitter currents $\vec{j}_\text{t}$, i.e. those entering the \ac{DMA} structure, but one can easily design for the currents supplied by the \ac{RF} chains $\vec{j}\in\mathbb{C}^{N\times 1}$, which are given by 
\begin{equation}
    \vec{j} = (\mat{i}_N + \mat{\Gamma})^{-1} \vec{j}_\text{t} = \mat{T}^{-1} \vec{j}_\text{t}, \label{eq:J_rfchains}
\end{equation}
where $\mat{T}$ is commonly known as the transmission coefficient \cite[eq. (2.51)]{Pozar2012}.

\subsection{Limitations imposed in phase-control by passive operation}
\label{subsec:Limitations}

When the imposed wave propagates along the waveguide, the \ac{DMA} elements sample the wave, apply a phase- and amplitude-shift dependent on the terminating admittance, and scatters the wave into free space and back into the waveguide. This, together with digital preprocessing at the \ac{RF} chains, is the basic mechanism through which \ac{DMA} can be useful to implement beamforming techniques. However, due to the passive operation of the \ac{DMA} element terminations, the amplitude and phase of the scattered wave are related and usually constrained to a subset of the complex plane (e.g., through Lorentzian constraint \cite{Smith2017, Schlezinger2019}). To illustrate this, consider a toy example with a \ac{DMA} comprising of a single waveguide ($N=1$) with a single \ac{DMA} element ($L=1$) on the upper boundary transmitting to a single receiver ($M=1$). Then, \eqref{eq:IO_FF} becomes
\begin{align}
    j_\text{r} = \left(Y_\text{r}+Y_\text{rr}\right)^{-1} \left(Y_\text{rs}\vartheta Y_\text{st}\right)j_\text{t}. 
\end{align}
where $\vartheta = \left(Y_\text{s}+Y_\text{ss}\right)^{-1}$ is the operation done by the \ac{DMA} element before scattering the wave towards the receiver.
The power dissipated in the \ac{DMA} element attached admittance $Y_\text{s}$ (i.e., the admittance that is used to tune the element) is given as
\begin{align}
    P_d = \frac{\Re{-j_\text{s}^* v_\text{s}}}{2} \overset{(a)}{=}\frac{\|j_\text{s}\|^2\Re{ Y_\text{s}}}{2} \label{eq:powerDissipatedDMA}
\end{align}
where \eqref{eq:ohm1} is used in $(a)$. 

\begin{figure}[t]
    \centering
    \includegraphics[width = \columnwidth]{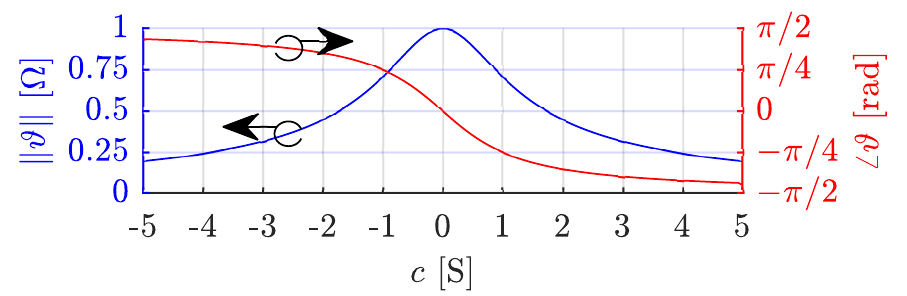}
    \caption{Dependency between amplitude and phase of the wave scattered by a passive \ac{DMA} element for $\Re{Y_\text{ss}} = 1$.}
    \label{fig:ampPhaseDependency}
\end{figure}

As shown in \eqref{eq:powerDissipatedDMA}, if \mbox{$\Re{Y_\text{s}} < 0$} the dissipated power becomes negative, meaning that the termination is injecting power into the system and thus contradicting the passive behaviour assumed for the \ac{DMA} elements. When $\Re{Y_\text{s}} = 0$, the termination is lossless and scatters the incoming wave without dissipating any power. For simplicity, let us stick to the latter case, and consider $Y_\text{s} = i\left(c - \Im{Y_\text{ss}}\right)$ where $c\in\mathbb{R}$ is an arbitrary constant.

The operation done by the \ac{DMA} element is then $\vartheta = \left(\Re{Y_\text{ss}} + i \, c \right)^{-1}$, with phase and amplitude given by
\begin{align}
    \angle \vartheta &= -\atan\left(\frac{c}{\Re{Y_\text{ss}}}\right) \phantom{a},\phantom{a} \Re{Y_\text{ss}} > 0, \label{eq:Response_angle}\\
   \|\vartheta \| &= \frac{1}{\sqrt{c^2 + \Re{Y_\text{ss}}^2}}. \label{eq:Response_mod}
\end{align}
The domain of $\angle \vartheta$ is hence limited to $[-\frac{\pi}{2},\frac{\pi}{2}]$ for an arbitrary $c \in \mathbb{R}$ and $\Re{Y_\text{ss}} > 0$ (which is also imposed by the passiveness of the elements). Notably, the amplitude of the scattered wave is not constant for the whole phase range, being the maximum at $\angle \vartheta = 0$. As the phase approaches $-\frac{\pi}{2}$ or $\frac{\pi}{2}$, the amplitude approaches zero, meaning that the effective range of phases is even smaller. Fig. \ref{fig:ampPhaseDependency} shows the trend of $\vartheta$ for $\Re{Y_\text{ss}} = 1$. Note that, performing basic algebraic manipulations, the response in \eqref{eq:Response_angle}-\eqref{eq:Response_mod} is proved to be equivalent to the so-called Lorentzian response in, e.g., \cite[Eq. (3)]{Smith2017}. Therefore, the amplitude and phase of the wave scattered by the passively operated \ac{DMA} element is tightly linked due to the restriction of $\Re{Y_\text{ss}} \geq 0$. In other words, the equivalent impedance resulting from the \ac{DMA} element operation over the incoming signal is restricted to the positive complex semiplane. Similar results have been obtained recently in \cite{Qian2021} in the context of reflecting surfaces.

\section{DMA Analysis: Calculating $\mat{Y}_\text{st}$, $\mat{Y}_\text{tt}$ and $\mat{Y}_\text{ss}$ }\label{sec:DMAanalysis}
As shown before, the input-output relationship and the performance of \ac{DMA} based systems depend on the value of the different admittance matrices. In the following sections, we derive tractable and closed-form expressions for all the submatrices in \eqref{eq:CircuitalModel} from the underlying electromagnetic theory, which is based on the magnetic dipole model in \cite{Pulido2017, Smith2017, Pulido2018}. 

\subsection{Waveguide analysis and antenna modeling}

We consider a rather generic model for the linear waveguides composing the \ac{DMA}, in which they are characterized as rectangular tubes of lossless dielectric in a planar slab of \ac{PEC} aligned with the $xz$-plane (see Fig. \ref{fig:system}). The \ac{PEC} slab is assumed infinite along the $x$ and $z$ dimensions. The rectangular tubes making out the waveguides have a width of $a$ along the $z$ direction, a height of $b$ along the $y$ direction, and a length of $S_\mu$ along the $x$ direction. The values of $a$ and $b$ are chosen such that the waveguide only supports propagation in the fundamental mode $\text{TE}_{10}$ around the frequency of interest.

The \ac{DMA} elements are modeled as sub-wavelength slots in the upper surface of the waveguide ($y=b$) which scatter (i.e., radiate) the incoming field out into free space and back into the waveguide. Through the surface equivalence theorem (Love's principle) \cite[eq. (7-43)]{Balanis2012}, the fields scattered by the \ac{DMA} elements can be determined based on an effective magnetic surface current ($\vec{s}_\text{m}$) and an effective electric surface current ($\vec{s}_\text{e}$) as\footnote{A reader familiar with electromagnetism may note a change with respect to the conventional notation. However, we deliberately denote the surface currents as $\vec{s}$ in order to clearly differentiate them from the magnetic currents flowing through the different ports in \eqref{eq:CircuitalModel}.}
\begin{align}
    \vec{s}_\text{m} &= -\hat{\vec{n}} \times \vec{e}_\text{s}, \label{eq:equivalentSurfaceCurrent} \\
    \vec{s}_\text{e} &= \hat{\vec{n}} \times \vec{h}_\text{s},
\end{align}
where $\hat{\vec{n}}$ is a unitary vector normal to the surface (in our case, $\hat{\vec{n}} =\pm \hat{\vec{y}}$ depending on whether the objective is to compute the field outside or inside the waveguide); and $\vec{e}_\text{s}$ and $\vec{h}_\text{s}$ are respectively the electric and magnetic fields in the \ac{DMA} element aperture (the boundary between the inside of the waveguide and free space). Given the effective surface currents, the electric and magnetic fields scattered by the \ac{DMA} element are given by \cite[Eqs. (4.181) and (4.184)]{Tai1994}
\begin{align}
    \vec{E}(\vec{r}) =& - i\omega\mu \oiint_S \mat{G}_\text{e1}\left(\vec{r}, \vec{r}'\right) \vec{s}_\text{e}(\vec{r}') \dif S' \notag \\
    &- \oiint_S \nabla \times \mat{G}_\text{e2}\left(\vec{r}, \vec{r}'\right) \vec{s}_\text{m}(\vec{r}') \dif S', \label{eq:EIntegral}\\
    \vec{H}(\vec{r}) =& \oiint_S \nabla \times \mat{G}_{\text{e}1}\left(\vec{r}, \vec{r}'\right) \vec{s}_\text{e}(\vec{r}') \dif S' \notag \\
    &-i \omega \epsilon \oiint_S \mat{G}_{\text{e}2}\left(\vec{r}, \vec{r}'\right) \vec{s}_\text{m}(\vec{r}') \dif S', \label{eq:HIntegral}
\end{align}
where $\omega$ is the angular frequency, $\mu$ and $\epsilon$ are the permeability and permittivity of the medium, $\vec{r}\in\mathbb{R}^3$ is an arbitrary point in the space, $S$ is the boundary surface, and $\mat{G}_\text{e1}\left(\vec{r}, \vec{r}'\right)\in\mathbb{C}^{3\times 3}$ and $\mat{G}_\text{e2}\left(\vec{r}, \vec{r}'\right)\in\mathbb{C}^{3\times 3}$ are the electrical Green's functions satisfying the Dirichlet and Neumann conditions respectively on the boundary of the domain, i.e.,
\begin{align}
    \hat{\vec{n}}\times\mat{G}_\text{e1}\left(\vec{r}, \vec{r}'\right) = 0,  \\
    \hat{\vec{n}}\times\nabla\times\mat{G}_\text{e2}\left(\vec{r}, \vec{r}'\right) = 0.
\end{align}

Note that, from a signal processing viewpoint, $\mat{G}_\text{e1}\left(\cdot\right)$ and $\mat{G}_\text{e2}\left(\cdot\right)$ can be seen as the 3D space impulse response of the propagation medium, relating the equivalent surface currents with the generated fields at any point in space.  

Assuming the \ac{DMA} elements are small compared to the wavelength, the Green's function is approximately constant over the surface of the element, and \eqref{eq:HIntegral} is rewritten as\footnote{The same can be done with \eqref{eq:EIntegral}.}
\begin{align}
    \vec{H}(\vec{r}) \approx  & -i \omega\nabla \times \mat{G}_{\text{e}1}\left(\vec{r}, \vec{r}'\right)  \vec{p} \notag \\
    &- \omega^2 \epsilon\mu \mat{G}_{\text{e}2}\left(\vec{r}, \vec{r}'\right)  \vec{m}, \label{eq:hfieldMoment}
\end{align}
where $\vec{p}$ and $\vec{m}$ are the electric and magnetic dipole moments of three ideal electric and magnetic dipoles oriented in the $x$, $y$ and $z$ directions, which are given by \cite[Sec. 9.5]{Jackson1999}
\begin{align}
    \vec{p}&=\frac{-1}{i \omega}\oiint_S \vec{s}_\text{e}(\vec{r}') \dif S',  \label{eq:elDipoleMoment}\\
    \vec{m}&= \frac{-1}{i\omega\mu}\oiint_S \vec{s}_\text{m}(\vec{r}') \dif S'. \label{eq:magDipoleMoment}
\end{align}

For sub-wavelength apertures, the electric and magnetic moments are proportional respectively to the electric and magnetic incident fields in the \ac{DMA} scattering slot, i.e., from \cite[Eq. (9.73)]{Jackson1999} we have\review{
\begin{align}
    \vec{p} &= \epsilon a_e \vec{e}_0(\vec{r}), \\
    \vec{m} &=  \mat{A}_m \vec{h}_0(\vec{r}),
\end{align}}
with $a_e\in\mathbb{C}$ and $\mat{A}_m\in\mathbb{C}^{3\times 3}$ being the effective electric and magnetic polarizabilities, and $\vec{e}_0$ and $\vec{h}_0$ being the incident\footnote{With the slot in place, the total field is given by the sum of the incident field and the field scattered by the slot itself, i.e., $\vec{e}_0(\vec{r}) + \vec{e}(\vec{r})$.} wave at the position of the ideal dipole. In general, one can obtain the effective polarizability corresponding to some specific scattering element through e.g. direct evaluation of \eqref{eq:magDipoleMoment} and \eqref{eq:elDipoleMoment} using simulated field distributions or using a phase-corrected S-parameter measurement as in \cite{karamanos_polarizability_2012,Pulido2017}. By this approach the model can be fitted to the exact \ac{DMA} element design of interest. However, the objective of this work is the analysis and design of generic \ac{DMA} systems, so the effective polarizability is approximated to capture overall trend of common designs and not to fit a specific element design. 

Assuming that resonant length of the \ac{DMA} elements are aligned with the axis of the waveguide, then $\mat{A}_m$ can be assumed diagonal and thus the magnetic dipole moment has the same non-zero components as the incident field. Secondly, the \ac{PEC} boundary conditions on the waveguide walls require that $\nvec{n} \cross \vec{e} = \vec{0}$ and $\nvec{n}^T \vec{h} = 0$, where $\nvec{n}$ is the unit normal vector of the waveguide wall. This means that only the $y$-component of $\vec{p}$ can be excited and only the $x$ and $z$ components of $\vec{m}$ can be excited. Moreover, as shown in \cite{Pulido2017}, the electric dipole moment $\vec{p}$ is negligible for common resonant \ac{DMA} element designs on the waveguide boundary, and hence it completely vanishes from \eqref{eq:hfieldMoment}. Finally, since we are choosing the waveguide dimensions such that only the fundamental $\text{TE}_{10}$ mode propagates, the magnetic field along the center of the waveguide (i.e., $z=a/2$) only has a non-zero component in the $z$ axis \cite[Sec. 3.3]{Pozar2012}. Putting all the pieces together, we have that the field scattered by a small resonant \ac{DMA} element can be accurately approximated around the resonance frequency as the field radiated by an infinitesimal magnetic dipole oriented in the $z$ direction, which carries a current given as
\begin{equation}
    \vec{s}_\text{m}(\vec{r}) = s_m \delta(x - x')\delta(z - z')\nvec{z}, \label{eq:SourceCurrent}
\end{equation}
where $\vec{r}' = (x', y', z')$ is the position of the dipole, $s_m\in\mathbb{C}$ is the magnitude of the current and $\delta(\cdot)$ is the delta function. 

\subsection{General procedure to derive mutual admittances with the infinitesimal dipole model}

With the input-output relationships derived in terms of the mutual admittances in Section \ref{sec:InsertionLosses}, and the inherited magnetic dipole model from \cite{Pulido2017, Smith2017} adapted in the previous subsection, the remaining step is deriving expressions for all the admittance matrices in \eqref{eq:CircuitalModel}. To that end, we apply the following three-steps procedure, which relies on the \ac{CMS} antenna assumption, to characterize the mutual admittance between two agents in the system:
\begin{enumerate}
    \item One agent --- equivalently, one port in Fig. \ref{fig:network} --- is excited with a magnetic source current $\vec{s}_\text{m}$ as in \eqref{eq:SourceCurrent}. Then, introducing \eqref{eq:SourceCurrent} in \eqref{eq:HIntegral} and applying the infinitesimal magnetic dipole moment, we obtain
    \begin{equation}
    \vec{h}(\vec{r}) = - i s_m \omega \epsilon  \mat{G}_{\text{e}2}\left(\vec{r}, \vec{r}'\right) \nvec{z}. \label{eq:h}
    \end{equation}
    \item The induced magnetic voltage in the receiver port is calculated as the integral of the field over the magnetic dipole, i.e., $v_m = -\int_0^l \nvec{z}^T\vec{h}(\vec{r})\dif z$ \cite[Eq. (38)]{Tkatchenko1995}\cite[Eq. (7)]{Agrawal1980}. Assuming infinitesimal dipoles and considering $l=1$ without any loss of generality, we have
    \begin{equation}
        v_m = - \nvec{z}^T \vec{h}\p{\vec{r}} = i\omega \epsilon s_m {G}_{\text{e}2,zz}\left(\vec{r}, \vec{r}'\right), \label{eq:VmGeneral}
    \end{equation}
    where ${G}_{\text{e}2,zz} = (\mat{G}_\text{e2})_{3,3}$, which relates the $z$ component of the field at a point $\vec{r}$ with the $z$ component of the source distribution at a point $\vec{r}'$. Note that, under the infinitesimal dipole model, $l$ is just a scaling factor which cancels under transmit power normalization, and therefore can be omitted without compromising the validity of the model proposed here.
    \item The mutual admittance between source port $l$ and receiver port $n$ is determined as
    \begin{equation}
        (\mat{Y})_{l,n} = \frac{v_m}{s_m} = i\omega \epsilon  {G}_{\text{e}2,zz}\left(\vec{r}_l, \vec{r}_n\right). \label{eq:Ygeneral}
    \end{equation}
\end{enumerate}

Using this procedure, the coupling between the ports in the network is derived in the following sections, starting with the coupling between the transmitters and \ac{DMA} elements.

\subsection{Coupling between transmitters and \ac{DMA} elements --- $\mat{Y}_\text{st}$}

To characterize $\mat{Y}_\text{st}$, we consider the distance from the source (i.e., the input to the waveguide where the \ac{RF} chain is connected) to the nearest \ac{DMA} element is large enough such that the source distribution does not impact the received field at the first element in the waveguide. A distance above one wavelength is sufficient for this, and hence this is but a minor assumption. However, it allows us to choose an arbitrary source distribution. Therefore, for simplicity, we model the source identically to the \ac{DMA} elements with a current given by \eqref{eq:SourceCurrent}, and thus from \eqref{eq:VmGeneral} we have
\begin{equation}
    v_m = i\omega \epsilon s_m \nvec{z}^T \mat{G}^{(w)}_{\text{e}2}\left(\vec{r}, \vec{r}'\right)\nvec{z} = i\omega \epsilon s_m {G}^{(w)}_{\text{e}2,zz}\left(\vec{r}, \vec{r}'\right), \label{eq:vmYst}
\end{equation}
where ${G}^{(w)}_{\text{e}2,zz} = (\mat{G}^{(w)}_{\text{e}2})_{3,3}$ is the third diagonal component of the Green's function inside the waveguide. Specifically, ${G}^{(w)}_{\text{e}2,zz}$ is given by (see appendix \ref{app:A1} for additional details)
\begin{align}
&G_{e2,zz}^{(w)}(\vec{r},\vec{r}')=\frac{-k_x\es{\frac{\pi z}{a}} \es{\frac{\pi z'}{a}} }{ a b k^2} \notag \\
&\times\frac{\ec{k_x \p{x'+x-S_\mu}} + \ec{k_x\p{S_\mu-|x-x'|}}}{\es{k_x S_\mu}}.\label{eq:Gw_xx_param}
\end{align}

In \eqref{eq:Gw_xx_param},  $0 \leq x,x'\leq S_\mu$, $0\leq y,y' \leq b$, and $0\leq z,z' \leq a$; and $k_x$ is defined according to \cite[Eq. (8-18)]{Balanis2012} as
\begin{align}
    k_x = \Re{\sqrt{k^2-(\pi/a)^2}} - i \Im{\sqrt{k^2-(\pi/a)^2}},
\end{align}
with $k^2 = (\pi/a)^2 + k_x^2 = \omega^2\epsilon\mu$. Note that, for the propagating $\text{TE}_{10}$ mode, $k_x\in\mathbb{R}$. Finally, the elements of $\mat{Y}_\text{st}$ are easily computed as 
\begin{equation}
    (\mat{Y}_\text{st})_{l,n} = \frac{v_m}{s_m} = i\omega \epsilon {G}^{(w)}_{\text{e}2,zz}\left(\vec{r}_l, \vec{r}_n\right), \label{eq:Yst}
\end{equation}
where $\vec{r}_l$ for $l= 1,\dots,L$ is the position of the $l$-th radiating element and $\vec{r}_n$ is the position vector of the $n$-th \ac{RF} chain. As a remark, note that \eqref{eq:Yst} is valid only for \ac{RF} chains and \ac{DMA} elements placed on the \textit{same} waveguide. In turn, we have that $(\mat{Y}_\text{st})_{l,n} = 0$ for each pair $l,n$ placed in different waveguides (valid for any \ac{CMS} antenna \cite{Wasylkiwskyj1970}, e.g., magnetic dipoles, since they become electrically invisible when shorted and thus no field scatters). 

\subsection{Transmitters self admittances --- $\mat{Y}_\text{tt}$}

With $\mat{Y}_\text{st}$, the calculation of the self-admittance matrix $\mat{Y}_\text{tt}$ is straightforward.  Let us impose the very mild assumption in which the distinct \ac{RF} chains are isolated between them, implying that $\mat{Y}_\text{tt}$ is a diagonal matrix. Then, under the infinitesimal magnetic dipole model, the self-admittance of the transmitters is directly given by
\begin{equation}
    (\mat{Y}_\text{tt})_{n,n} = \frac{-2 i k_x \es{\frac{\pi}{a}z_n'}^2\ec{k_x S_\mu}}{ab  \omega \mu \es{k_x S_\mu}},\label{eq:Ytt}
\end{equation}
which is easily proved by introducing \eqref{eq:Gw_xx_param} in \eqref{eq:Yst} and particularizing for $\vec{r} = \vec{r}'$, being $z_n'\in (0,a)$ the $z$ coordinate of the $n$th \ac{RF} chain. 

\subsection{Coupling between DMA elements --- $\mat{Y}_\text{ss}$} \label{sec:couplingDMAElements}

The last effect to be characterized in the \ac{DMA} is the mutual coupling between radiating elements. This coupling depends on whether two radiating elements are positioned on the same waveguide or not.  In the former case, the interaction between the two elements occurs  both through the waveguide and the air, whilst in the latter case the interaction is only through the air. Hence, in general, the elements $(\mat{Y}_\text{ss})_{l,l'}$ for $l,l' = 1,\dots,L$ are given by the ratio between the voltage in the terminals of the $l$th element due to the current applied in the $l'$th element, as in previous cases. Following the same previous steps, we obtain
\begin{equation}
    (\mat{Y}_\text{ss})_{l,l'} = i\omega \epsilon {G}_{\text{e}2,zz}\left(\vec{r}_l, \vec{r}_{l'}\right). \label{eq:YssGeneral}
\end{equation}

As stated before, depending on whether the $l$th and $l'$th elements are placed on the same waveguide or not, we introduce the corresponding Green's function in \eqref{eq:YssGeneral}. Moreover, according to the system model previously introduced, the waveguides and hence the \ac{DMA} elements are placed on top of an infinite \ac{PEC} plane. This implies that, due to image theory \cite[sec. 12.2]{Balanis2016}, the effective source current is doubled. Therefore, we have that
\begin{align}
    {G}_{\text{e}2,zz}&\left(\vec{r}_l, \vec{r}_{l'}\right) = 2{G}_{\text{e}2,zz}^{(a)}\left(\vec{r}_l, \vec{r}_{l'}\right) +\notag \\
    & \begin{cases}
    {G}_{\text{e}2,zz}^{(w)}\left(\vec{r}_l, \vec{r}_{l'}\right) , & \text{\footnotesize for $l$, $l'$ in same waveguide}\\
    0, & \text{\footnotesize for $l$, $l'$ in different waveguides}
    \end{cases}, \label{eq:Ge2xx_generic}
\end{align}
where ${G}_{\text{e}2,zz}^{(w)}$ is given in \eqref{eq:Gw_xx_param} and ${G}_{\text{e}2,zz}^{(a)}$ is the third diagonal component of the Green's function in free space given by (see Appendix \ref{app:A2})
\begin{align}
    {G}^{(a)}_{\text{e}2,zz}\left(\vec{r},\vec{r}'\right) =&  \left( \frac{R^2 - \Delta z^2}{R^2} - i\frac{ R^2-3\Delta z^2}{R^3 k}\right. \notag \\
    &\left.-\frac{ R^2-3\Delta z^2}{R^4 k^2}\right) \frac{\e{-i k R}}{4 \pi R}, \label{eq:GfreeSpaceFull}
\end{align}
with $R = \|\vec{r}-\vec{r}'\|_2$ and $\Delta z =  z-z'$.

Finally, the self impedance of the \ac{DMA} elements is obtained as 
\begin{align}
    (\mat{Y}_\text{ss})_{l,l} = \lim_{\vec{r}_l\rightarrow \vec{r}_{l'}}i\omega \epsilon \left({G}^{(w)}_{\text{e}2,zz}\left(\vec{r}_l, \vec{r}_{l'}\right)+2{G}^{(a)}_{\text{e}2,zz}\left(\vec{r}_l, \vec{r}_{l'}\right)\right). \label{eq:YssSelf}
\end{align}

As a consequence of the infinitesimal magnetic dipole model we are using, it can be proved that $\Im{(\mat{Y}_\text{ss})_{l,l}}$ diverges due to the singularity of $G^{(a)}_{\text{e}2,zz}\left(\vec{r}_l, \vec{r}_{l'}\right)$ when $\vec{r}_l\rightarrow \vec{r}_{l'}$. This is not but an artifact of the theoretical model used here --- any real antenna will have a finite admittance --- which does not come at the price of any loss of generality. To circumvent this issue, we introduce a term $i\overline{Y}_\text{ss} = i \Im{\review{G^{(a)}_{\text{e}2,zz}\left(\vec{r}_l, \vec{r}_{l}\right)}}$ in the tunable admittances $\mat{Y}_\text{s}$ such that, in practice, one would have $\mat{Y}_\text{s}^{(\text{eq})} = \mat{Y}_\text{s} - i \overline{Y}_\text{ss}\mat{I}_L$. For the sake of simplicity, we here just neglect the imaginary part of the self admittance \review{from the air interface}, since our goal is providing a general framework for the analysis of \ac{DMA} systems. Therefore, we have
\begin{align}
    (\mat{Y}_\text{ss})_{l,l} =& \lim_{\vec{r}_l\rightarrow \vec{r}_{l'}}i\omega \epsilon \left({G}^{(w)}_{\text{e}2,zz}\left(\vec{r}_l, \vec{r}_{l'}\right)\right. \notag \\
    &\left.+2\Re{{G}^{(a)}_{\text{e}2,zz}\left(\vec{r}_l, \vec{r}_{l'}\right)}\right), \label{eq:YssSelf}
\end{align}
which, as shown in Appendix \ref{app:A3}, leads to
\begin{align}
    (\mat{Y}_\text{ss})_{l,l} =& \frac{k\omega \epsilon}{3\pi} - \notag \\
    &\frac{ i k_x \p{\ec{k_z\p{S_\mu -2 x'}}-\ec{k_x S_\mu}}}{ab  \omega \mu \es{k_x S_\mu}}. \label{eq:Yss_self}
\end{align}

\section{Coupling between Users and Wireless Channel: Calculating $\mat{Y}_\text{rr}$ and $\mat{Y}_\text{rs}$}\label{sec:UserCoupling}
With the analysis of the \ac{DMA} structure accomplished and closed formed expressions for the matrices $\mat{Y}_\text{st}$, $\mat{Y}_\text{tt}$ and $\mat{Y}_\text{ss}$, the next step is the characterization of both the coupling between the users $\mat{Y}_\text{rr}$ and the wireless channel $\mat{Y}_\text{rs}$. This analysis is carried out in the sequel. 

\subsection{Coupling between users --- $\mat{Y}_\text{rr}$}

Although in most cases the coupling between the users is negligible since they are sufficiently separated from each others (i.e., $\mat{Y}_\text{rr}$ is a diagonal matrix), for the sake of completeness in the model we here characterize the coupling between them as well as their self-impedance. 

By also modelling the users as ideal magnetic dipoles and following similar steps as in the previous section, we have that the magnetic voltage between the terminals of an arbitrary user at position $\vec{r}_m$ due to the impulsive magnetic current of magnitude $s_m$ in another user at $\vec{r}_{m'}$ is given as
\begin{equation}
    v_m = i\omega \epsilon s_m{G}^{(a)}_{\text{e}2,zz}\left(\vec{r}_{m}, \vec{r}_{m'}\right), \label{eq:vmYrr}
\end{equation}
with ${G}^{(a)}_{\text{e}2,zz}$ as in \eqref{eq:GfreeSpaceFull}. Hence, the elements of the mutual admittance matrix are directly given by
\begin{equation}
    (\mat{Y}_\text{rr})_{m,m'} = \frac{v_m}{s_m} = i\omega \epsilon {G}^{(a)}_{\text{e}2,zz}\left(\vec{r}_m, \vec{r}_{m'}\right). \label{eq:Yrr}
\end{equation}
The self admittance of the users, i.e., the diagonal elements of $\mat{Y}_\text{rr}$, is given as in the case of $\mat{Y}_\text{ss}$ as 
\begin{equation}
    (\mat{Y}_\text{rr})_{m,m}  = \lim_{\vec{r}_m\rightarrow \vec{r}_{m'}} i\omega \epsilon {G}^{(a)}_{\text{e}2,zz}\left(\vec{r}_m, \vec{r}_{m'}\right), 
\end{equation}
whose imaginary part diverges again. We apply the same assumption and consider that the equivalent load admittance at the users accounts for the imaginary part of the self admittance, i.e., $\mat{Y}_\text{r}^{(\text{eq})} = \mat{Y}_\text{r} - i \Im{(\mat{Y}_\text{rr})_{m,m}}\mat{I}$. Hence, with a slight abuse of notation, we have 
\begin{align}
    (\mat{Y}_\text{rr})_{m,m} &= \Re{ \lim_{\vec{r}_m\rightarrow \vec{r}_m'}i\omega \epsilon \p{{G}^{(a)}_{\text{e}2,zz}\left(\vec{r}_m, \vec{r}_{m'}\right)}} =\frac{k\omega \epsilon}{6\pi}. \label{eq:YrrSelf}
\end{align}
Note that \eqref{eq:YrrSelf} is exactly one half of the real part of \eqref{eq:Yss_self}, since in this case the dipoles are placed in free space instead of on top of a \ac{PEC} plane.

 \begin{figure}[t]
     \centering
     \includegraphics[width=0.6\linewidth]{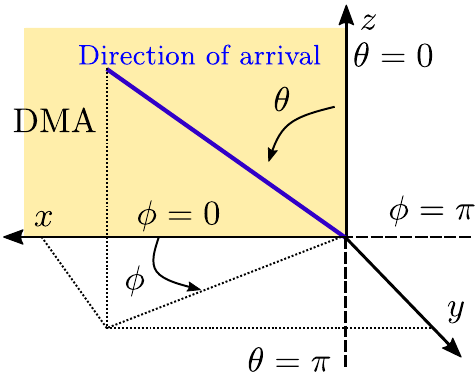}
     \caption{Definition of spherical coordinates with respect to Cartesian ones. The domain of the polar angle is $\theta \in \mathbb{R} : 0 \leq \theta \leq \pi$ and the domain of the azimuth angle $\phi \in \mathbb{R} : 0 \leq \phi < 2\pi$.}
     \label{fig:Coordinates}
 \end{figure}

\subsection{Wireless propagation channel --- $\mat{Y}_\text{rs}$}

As introduced before, the matrix $\mat{Y}_\text{rs}$ represent the wireless channel between the users and the reconfigurable elements in the \ac{DMA}, hence capturing the impact of the propagation medium. In general, the elements of $\mat{Y}_\text{rs}$ will depend on the characteristics of the environment, and then they can be given by deterministic models, stochastic models or empirical and simulated values. In this work, we give expressions for two commonly used channels in the context of \ac{LIS} and \ac{mMIMO}, namely deterministic \ac{LoS} model (or free space propagation) and spatially correlated Rayleigh fading, which are adapted to account for the particularities of the generic \ac{DMA} communication model. \\
\subsubsection{Deterministic channel}

The first considered channel is the so-called \ac{LoS} channel, where the entries of $\mat{Y}_\text{rs}$ are deterministic and dependent on the distance between transmitter and receiver. Thus, this model accounts only for path-loss and neglects the impact of the wireless propagation environment and hence the fading. It is used for high frequency systems where the reflected waves arrive at the receiver with negligible power due to the high pathloss exponent. Recently, it has also been used when analyzing theoretical properties of \ac{LIS} based systems \cite{Dardari2020, Hu2018}. 

In this case, the elements of $\mat{Y}_\text{rs}$ are just the mutual admittances between each antenna element in the \ac{DMA} and the users, which from \eqref{eq:Yrr} and the image theory reads
\begin{equation}
    (\mat{Y}_\text{rs})_{l,m} = -i\omega \epsilon 2{G}^{(a)}_{\text{e}2,zz}\left(\vec{r}_l, \vec{r}_{m}\right), \label{eq:Yrs_los}
\end{equation}
with $\vec{r}_l$ for $l= 1,\dots,L$ and $\vec{r}_m$ for $m=1,\dots,M$ are respectively the position vector for the \ac{DMA} elements and the users. In \eqref{eq:Yrs_los}, the negative sign is due to the sign-change of the normal vector in \eqref{eq:equivalentSurfaceCurrent}. As the equivalent surface current $\vec{s}_m$ is determined with $\nvec{n}$ pointing into the waveguide, when determining the field outside the waveguide, the effective surface current is equal to $-2\vec{s}_m$.

If the users are positioned in the far-field, i.e., $R$ is very large in \eqref{eq:GfreeSpaceFull}, then the terms depending on $R^3$ and $R^4$ can be neglected and we obtain
\begin{equation}
    \left.{G}^{(a)}_{\text{e}2,zz}\left(\vec{r},\vec{r}'\right)\right|_{R\Uparrow} = \frac{\e{-i k R}}{4 \pi R} \es{\theta}^2, \label{eq:GfreeSpaceFF}
\end{equation}
with the polar and azimuth angles defined as in Fig. \ref{fig:Coordinates}. Introducing \eqref{eq:GfreeSpaceFF} in \eqref{eq:Yrs_los} renders the expression for the far field \ac{LoS} channel as
\begin{equation}
    (\mat{Y}_\text{rs})_{m,l} = -i\omega \epsilon 2\frac{\e{-i k R}}{4 \pi R} \es{\theta}^2, \label{eq:LoSChannel}
\end{equation}
Note that \eqref{eq:LoSChannel} corresponds to the widely-used Friis' propagation formula, being the differences the introduction of the radiation pattern of both the transmitter and receiver dipoles (represented by $\es{\theta}^2$) and the amplitude factor corresponding to the image theory.\\

\subsubsection{Stochastic channel model}
Although not as site-specific as empirical models, stochastic models are far more general and allow to extract overall trends and results \cite{Pizzo2020}. Due to the popularity of the correlated Rayleigh model in the \ac{mMIMO} literature, we particularize this model so that it is valid for the infinitesimal dipole characterization of \acp{DMA} we are performing. To that end, we rewrite the admittance matrix $\mat{Y}_\text{rs}$ as 
\begin{equation}
     \mat{Y}_\textrm{rs} = \begin{bmatrix}
     \vec{y}_1 & \vec{y}_2 & \cdots & \vec{y}_M
     \end{bmatrix}^T,
\end{equation}
where $\vec{y}_m \in\mathbb{C}^{L\times 1}$ for $m=1,\dots,M$ is the channel vector from $m$th user to the \ac{DMA}. Under far-field conditions --- i.e., same angle of departure for each radiating element in the \ac{DMA} --- the mutual admittance to the $m$th user can be written from \cite[eq. (6)]{Bjornson2021} as 
\begin{equation}
    \vec{y}_m = \sum_{n = 1}^{N_p} \frac{\alpha_{n}}{\sqrt{N_p}} \es{\theta_{n}}\es{\vartheta_{n}} \vec{a}(\theta_n,\phi_n), \label{eq:ym_Rays}
\end{equation}
where $N_p$ is the number of paths, $\alpha_{n}$ are independent and identically distributed complex random variables with zero mean modelling the pathloss, phase, and polarization-shift of the $n$th path, and the $\es{\theta}^2$ term of \eqref{eq:LoSChannel} is separated in terms of mutually independent polar angles of departure and arrival denoted $\theta_n$ and $\vartheta_n$ respectively. Finally, $\vec{a}(\theta_n,\phi_n)\in\mathbb{C}^{L\times 1}$ are the so-called steering vectors, given by
\begin{equation}
    \vec{a}(\theta_n,\phi_n) = \begin{bmatrix}
     \e{i\vec{k}(\theta_n,\phi_n)\vec{r}_{1}} & \cdots & \e{i\vec{k}(\theta_n,\phi_n)\vec{r}_{L}}
     \end{bmatrix}^T
\end{equation}
with $\vec{r}_{l}$ for $l = 1,\dots,L$ is the vector position of the $l$-th \ac{DMA} element and 
\begin{equation}
    \vec{k}(\theta,\phi) = k\begin{bmatrix} \es{\theta}\ec{\phi} & \es{\theta}\es{\phi} & \ec{\theta}
    \end{bmatrix}.
\end{equation}

When $N_p \rightarrow \infty$, then the central limit theorem holds in \eqref{eq:ym_Rays} and hence $\vec{y}_m\sim\mathcal{CN}_L(\mat{0},\mat{\Sigma}_m)$, i.e., $\vec{y}_m$ becomes a circularly symmetric Gaussian vector with zero mean and covariance matrix $\mat{\Sigma}_m$ given by
\begin{align}
    \mat{\Sigma}_m=\sigma_\alpha^2\mathbb{E}_{\theta, \phi}\left[\es{\theta}^2\vec{a}(\theta,\phi)\vec{a}(\theta,\phi)^H\right] \mathbb{E}_\vartheta\left[\es{\vartheta}^2\right], \label{eq:SigmaE}
\end{align}
where $\sigma_\alpha^2$ is the variance of the random variables $\alpha_n \forall \, n$, which is readily obtained from \eqref{eq:Yrs_los} as
\begin{equation}
    \sigma_\alpha^2 = \p{\frac{2\omega\epsilon}{4\pi R}}^2L_p,
\end{equation}
with $L_p$ the power losses due to polarization mismatch. 

Assuming an isotropic propagation environment, the different paths are uniformly distributed in front of the \ac{BS} (positive $y$-axis) and thus the distribution of $\theta$, $\vartheta$, and $\phi$ is characterized by
\begin{align}
    f(\theta,\phi) &= \frac{\es{\theta}}{2\pi}, \quad 0\leq \theta,\phi\leq \pi \label{eq:PDF_theta}\\ f(\vartheta) &= \frac{\es{\vartheta}}{2}, \quad 0\leq \vartheta\leq \pi. \label{eq:PDF_vartheta}
\end{align}

Introducing \eqref{eq:PDF_theta} and \eqref{eq:PDF_vartheta} in \eqref{eq:SigmaE}, the $(s,q)$-th element of $\mat{\Sigma}_m$ is calculated as (see appendix \ref{app:solveIntegral})
\begin{align}
    (\mat{\Sigma}_m)_{s,q} &= \frac{\sigma_\alpha^2 2}{3} \Int_0^\pi\Int_0^\pi\es{\theta}^3 \e{-i\vec{k}(\theta,\phi)(\vec{r}_s-\vec{r}_q)}\dif\theta\dif\phi \label{eq:Sigma}\\
    &= \frac{4\pi \sigma_\alpha^2}{3} \left[\p{1+\frac{3 \Delta z^2}{R^4 k^2} - \frac{1 + \Delta z^2 k^2}{R^2 k^2}}\frac{\es{k R}}{k R} \right.\notag \\
    &\phantom{=}+ \left.\p{\frac{1}{k R} - \frac{3 \Delta z^2}{k R^3}}\frac{\ec{k R}}{k R} \right]  , \label{eq:SigmaFinal} 
\end{align}
which is only valid for source fields on the surface of the \ac{DMA}. 

Interestingly, observe that, as shown in \eqref{eq:SigmaE}, the radiation pattern of the receiver antenna only provides a scaling of the covariance among the transmitters, and therefore has no effect of the spatial correlation on the transmitter side. Moreover, in striking contrast to the spatial correlation for isotropic antennas \cite[Eq. (10)]{Bjornson2021}, the covariance in \eqref{eq:SigmaFinal} differs depending on whether the separation is along the $z$ or $x$ axis. 

\section{Integration in MIMO communication model}\label{sec:MIMOmodel}
\begin{table*}[t]
    \renewcommand{\arraystretch}{2}
     \centering
       \caption{Overview of the circuital model for \ac{DMA} systems, where the theoretical expressions for the mutual admittances are summarized. ${G}^{(w)}_{e2,zz}$ and ${G}^{(a)}_{e2,zz}$ are given by \eqref{eq:Gw_xx_param}  and \eqref{eq:GfreeSpaceFull}, respectively.}
     \label{tab:admittanceOverview}
     \begin{tabular}{c| c | c}
        \textbf{Actors}  & \textbf{Analytical expression} & \textbf{Reference}\\ \hline \hline
        \multirow{2}{*}{Transmitter $\overset{\text{Waveguide}}{\longleftrightarrow}$ \ac{DMA} elements} &  $(\mat{Y}_\text{st})_{l,n} = i\omega \epsilon {G}^{(w)}_{\text{e}2,zz}\left(\vec{r}_l, \vec{r}_n\right)$ for $l,n$ in same waveguide &  \multirow{2}{*}{\eqref{eq:Yst}}\\ & $(\mat{Y}_\text{st})_{l,n} = 0$ for $l,n$ in different waveguides &  \\ 
        \hline 
        \multirow{2}{*}{Transmitter $\overset{\text{Self admittance}}{\longleftrightarrow}$ Transmitter} &  $ (\mat{Y}_\text{tt})_{n,n} = \frac{-2 i k_x \es{\frac{\pi}{a}z_n'}^2\ec{k_zS_\mu}}{ab\omega \mu \es{k_x S_\mu}}$ & \multirow{2}{*}{\eqref{eq:Ytt}}\\ 
        &  $(\mat{Y}_\text{tt})_{n,m} = 0$ for $n\neq m$  &  \\
        \hline
        \multirow{2}{*}{\ac{DMA} element $\overset{\text{Waveguide $+$ Air}}{\longleftrightarrow}$ \ac{DMA} element} &  $(\mat{Y}_\text{ss})_{l,l'} = i\omega \epsilon {G}_{\text{e}2,zz}\left(\vec{r}_l, \vec{r}_l'\right)$ for $l\neq l'$ with ${G}_{\text{e}2,zz}$ as in \eqref{eq:Ge2xx_generic}&  \eqref{eq:YssGeneral}\\ & $ (\mat{Y}_\text{ss})_{l,l} = {k\omega \epsilon}/(3\pi)- \frac{ i k_x \p{\ec{k_z\p{S_\mu -2 x'}}+\ec{k_x S_\mu}}}{ab  \omega \mu \es{k_x S_\mu}}$ & \eqref{eq:Yss_self} \\ 
        \hline
        \multirow{2}{*}{User $\overset{\text{ Air}}{\longleftrightarrow}$ User} &  $(\mat{Y}_\text{rr})_{m,m'} = i\omega \epsilon {G}^{(a)}_{\text{e}2,zz}\left(\vec{r}_m, \vec{r}_{m'}\right)$ for $m\neq m'$ &  \eqref{eq:Yrr}\\ & $ (\mat{Y}_\text{rr})_{m,m} = {k\omega \epsilon}/(6\pi)$ & \eqref{eq:YrrSelf} \\ 
        \hline
        \multirow{2}{*}{\ac{DMA} elements $\overset{\text{ Air}}{\longleftrightarrow}$ Users (Wireless channel)} &  $(\mat{Y}_\text{rs})_{m,l} = -i\omega \epsilon 2\frac{\e{-i k R}}{4 \pi R} \es{\theta}^2$ (Far field \ac{LoS} channel) &  \eqref{eq:LoSChannel}\\ & $ \mat{Y}_\textrm{rs} = [
     \vec{y}_1 \;\; \cdots\;\; \vec{y}_M]^T$ with $\vec{y}_m\sim\mathcal{CN}_L(\mat{0},\mat{\Sigma}_m)$ (Rayleigh model)& \eqref{eq:Sigma} \\ 
     \end{tabular}
 \end{table*}

The proposed circuital model accounts for the interactions between all the actors in the \ac{DMA} based system setup, and as shown in Section \ref{sec:InsertionLosses}, establishes a relationship between the magnetic currents delivered by the \ac{RF} chains and those at the users in terms of the mutual admittance matrices. However, from a signal processing viewpoint, it is convenient to rewrite the model as
\begin{equation}
    \vec{y} = \mat{H}_\text{eq}\mat{B}\vec{x} + \vec{n}, \label{eq:MIMOmodel}
\end{equation}
in which $\vec{y}\in\mathbb{C}^{M\times 1}$ is the received complex baseband signal at the users measured in volts, $\vec{x}\in\mathbb{C}^{M\times 1}$  is the transmitted vector of symbols, $\mat{B}\in\mathbb{C}^{N\times M}$ is the precoding matrix, $\mat{H}_\text{eq}\in\mathbb{C}^{M\times N}$ is the equivalent channel and $\vec{n}\in\mathbb{C}^{M\times 1}$ is the noise term. Hence, $\mat{B}\vec{x}$ is the resulting complex vector that is introduced at the \ac{RF} chains and $\mat{H}_\text{eq}$ models the equivalent channel between the transmitters and the users, which includes the \ac{DMA} architecture. 

By inspection, we observe that $\mat{B}\vec{x} = \vec{j}_\text{t}$, and therefore from \eqref{eq:IO_FF} one has\footnote{An equivalent expression without the far field assumption is obtained by introducing \eqref{eq:IO}.} 
\begin{align}
    \mat{H}_\text{eq} =  \left(\mat{Y}_\text{r}+\mat{Y}_\text{rr}\right)^{-1} \left(\mat{Y}_\text{rs}\left(\mat{Y}_\text{s}+\mat{Y}_\text{ss}\right)^{-1}\mat{Y}_\text{st}\right).  \label{eq:Heq}
\end{align}  
Introducing \eqref{eq:Heq} in \eqref{eq:MIMOmodel} yields
\begin{equation}
    \vec{y} = \left(\mat{Y}_\text{r}+\mat{Y}_\text{rr}\right)^{-1} \left(\mat{Y}_\text{rs}\left(\mat{Y}_\text{s}+\mat{Y}_\text{ss}\right)^{-1}\mat{Y}_\text{st}\right)\mat{B}\vec{x} + \vec{n}, \label{eq:MIMOmodel_param}
\end{equation}
which gives the relationship between the transmitted and received symbols in terms of the circuital model. In a similar way as in hybrid analog and digital architectures, the beamforming operation is performed in two steps: \textit{i)} the digital precoding in $\mat{B}$ and \textit{ii)} the tuning carried out at the \ac{DMA} elements, which is done by changing $\mat{Y}_\text{s}$ and is constrained for passive systems as detailed in Section \ref{subsec:Limitations}. From \eqref{eq:powerRecieved}, \eqref{eq:powerTransmitted}, and \eqref{eq:powerSupplied}, the received power in the $m$th user, the transmitted power, and the supplied power are expressed in terms of the complex symbols as 
\begin{align}
    P_{\text{r}m} &= \frac{1}{2} \Re{\p{\mat{Y}_{\text{r}}}_{m,m}} \|(\vec{y})_m\|^2,\\
    P_\text{t} &= \frac{1}{2}\Re{\vec{x}^H\mat{B}^H\mat{Y}_\text{p}\mat{B}\vec{x}}, \\
    P_\text{s} &= \frac{1}{2}\Re{\vec{x}^H\mat{B}^H\p{\mat{I}_N-\mat{\Gamma}^H\mat{\Gamma}}^{-1}\mat{Y}_\text{p}\mat{B}\vec{x}}, 
\end{align}
which, together with \eqref{eq:MIMOmodel_param}, are the starting point for transmission design algorithms. Then, all the electromagnetic theory is embedded in a digital communication model in the same form as those used in \ac{mMIMO}, and the beamforming design can be carried out in a similar way --- see, e.g., \cite{Joham2005}. The difference is that, compared to conventional \ac{mMIMO}, we have two optimization variables: the digital beamforming matrix $\mat{B}$ and the tunable admittances $\mat{y}_\text{s}$, exactly as in hybrid systems. Naturally, the domain of $\mat{y}_\text{s}$ would depend on the physical realization of the \ac{DMA}, and can either be set to discrete values for a digital implementation or could be controllable within some dynamic range for an analog implementation. However, in this work, we deliberately do not impose any restriction on the value of $\mat{y}_\text{s}$, aiming to provide a model as general as possible.

Finally, for reader's convenience, a summary of the admittance model is provided in Table \ref{tab:admittanceOverview}. Also, for the sake of reproducibility, the model has been implemented in \textsc{Matlab} and simulation code made available at \cite{DMACode2022}.

\section{Validation through Full-Wave Simulations}\label{sec:Simulations}
\begin{figure}[t]
     \centering
     \includegraphics[width=0.7\linewidth]{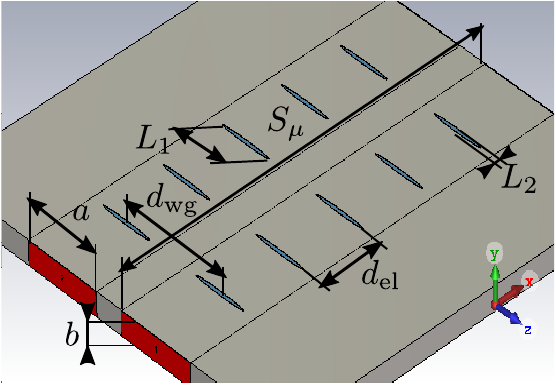}
     \caption{CST model of the two waveguide DMA with two elliptical elements. The elements are centered along the waveguides and spaced $d_\text{el}=0.6\lambda$. Parameters are given in Table \ref{tab:simValues}.}
     \label{fig:CSTModel}
 \end{figure}

To validate the modelling procedure, the predicted results by the model are compared with full-wave simulations conducted in CST Studio Suite. Specifically, we consider a simple \ac{DMA} composed by two waveguides with five radiating elements embedded in each of them. The two waveguides are placed in parallel in a slab of \ac{PEC}, with a center-to-center spacing of $d_\text{wg}$. Along the waveguides, five small cylindrical slots with major radius $L_1$ and minor radius $L_2$ are cut in the upper wall, which acts as equivalent magnetic dipoles \cite{Pulido2017} with termination admittances $\mat{Y}_\text{s} = \mat{i}_{10} \left( 2 - 15.7934i\right)$. Note that changing the size of the major and minor radii corresponds to changing the termination admittances $\mat{Y}_\text{s}$. Therefore, despite lacking any type of reconfigurability, the slots used here are representative of a \ac{DMA} with an arbitrary configuration (i.e., a given value of $\mat{Y}_\text{s}$), and are thus valid to check the proposed theoretical model. Moreover, the front-facing ends of the waveguides are terminated in waveguide-ports exciting the fundamental mode of the waveguide and thus representing the \ac{RF} chains. In turn, the back-facing ends are terminated in a plane of \ac{PEC} in order to account for the finite length of the waveguides and hence the reflections inside them. The complete simulated layout is depicted in Figure \ref{fig:CSTModel}, while all the parameters are summarized in Table \ref{tab:simValues}. As the waveguide port ensures that there is no reflection at the interface, the connector intrinsic impedance $Y_0$ has to be chosen to emulate this behavior for comparison. This is done by matching $Y_0$ to the self-admittance $\p{\mat{y}_\text{tt}}_{n,n}$ in a semi-infinite waveguide where the input is placed at the terminated end of the waveguide. Knowing the equivalent intrinsic impedance of the CST waveguide port $Y_0$, the effective termination of the DMA element can be extracted using the $S_{11}$-parameter of a single finite waveguide with single DMA element in the upper wall. In this case the $S_{11}$-parameter is equal the reflection coefficient in \eqref{eq:Gamma}, whereby the equivalent termination $Y_\text{s}$ can be isolated as the only unknown.

\begin{table}[t]
    \renewcommand{\arraystretch}{1.4}
     \centering
       \caption{Parameters used in CST simulation.}
       \label{tab:simValues}
  \begin{tabular}{c|c | c}
        \textbf{Parameter}  & \textbf{Symbol} & \textbf{Value}\\ \hline
        Frequency & $f$ & $\SI{10}{\giga\hertz}$\\
        Waveguide width  & $a$ & $0.7318\lambda$ \\
        Waveguide height  & $b$ & $0.1668\lambda$ \\
        Waveguide length  & $S_\mu$ & $\SI{110}{\milli\meter}$ \\
        Waveguide wall thickness  & $t$ & $\SI{0.1}{\milli\meter}$  \\
        Element major radius  & $L_1$ & $\SI{0.5}{}\lambda$ \\
        Element minor radius  & $L_2$ & $\SI{0.033}{}\lambda$ \\
        Element spacing & $d_\text{el}$  & $0.6\lambda$\\
        Waveguide spacing & $d_\text{wg}$ & $1\lambda$\\
        Element effective termination & $Y_\text{s}$  & $\SI{2 - 15.7934i}{\siemens}$ \\
        Connector intrinsic impedance & $Y_0$ & $\SI{35.3387}{\siemens}$
  \end{tabular}
\end{table}
 
For simplicity, the supplied power is set to $P_s = \SI{1}{\watt}$ and is distributed equally among the two waveguide-ports. The impinging currents, i.e., the currents introduced in the waveguides, are given a phase of $\angle (\vec{j})_{n} = 0$ for $n\in \{1,2\}$. The current entering the network is then given by \eqref{eq:J_rfchains} as $\vec{j}_\text{t} = \mat{T} \vec{j}$. Notice that, in this setup, we have fixed the supplied power and we calculate the equivalent currents $\vec{j}_\text{t}$ entering the network.  

Applying Eqs. \eqref{eq:IO_FF} through \eqref{eq:powerSupplied} with the expressions for the mutual impedances in sections \ref{sec:DMAanalysis} and \ref{sec:UserCoupling} (see Table \ref{tab:admittanceOverview} for a summary) yields $P_\text{t} = \SI{0.6077}{\watt}$, $\vec{j} = 0.1682 \begin{bmatrix} 1& 1 \end{bmatrix}^T$, $\vec{j}_\text{t} = 0.2266 + 0.0877i \begin{bmatrix} 1 & 1 \end{bmatrix}^T$, and the currents in the \ac{DMA} elements as
   \begin{align}
       \vec{j}_\text{s} = \begin{bsmallmatrix}    0.1459 + 0.0510i \\ -0.0732 - 0.0409i\\  0.0249 + 0.0336i \\ 0.0010 - 0.0276i \\ -0.0128 + 0.0255i \\ 0.1459 + 0.0510i \\    -0.0732 - 0.0409i \\   0.0249 + 0.0336i \\   0.0010 - 0.0276i \\ -0.0128 + 0.0255i \end{bsmallmatrix}  \quad , \quad \left|\vec{j}_\text{s}\right| = \begin{bsmallmatrix}       0.1546 \\   0.0838 \\   0.0418 \\   0.0276 \\    0.0285 \\   0.1546 \\       0.0838 \\      0.0418 \\      0.0276 \\     0.0285\end{bsmallmatrix}. \label{eq:simCurrents}
   \end{align}
where $(\vec{j}_\text{s})_n$ for $n\in \{1,5\}$ are the currents in elements along the first waveguide and $(\vec{j}_\text{s})_n$ for $n\in \{6,10\}$ are the currents in the elements along the second waveguide. 

From \eqref{eq:simCurrents}, it can be observed that the amplitudes of the currents flowing through the \ac{DMA} element decrease with  increasing index. This is a coherent result, since as the signal is propagated through the waveguide, part of its power is radiated into free space by the previous elements, leaving less power left to the remaining elements along the waveguide.

\begin{figure}[t]
     \centering
     \includegraphics[width=\linewidth]{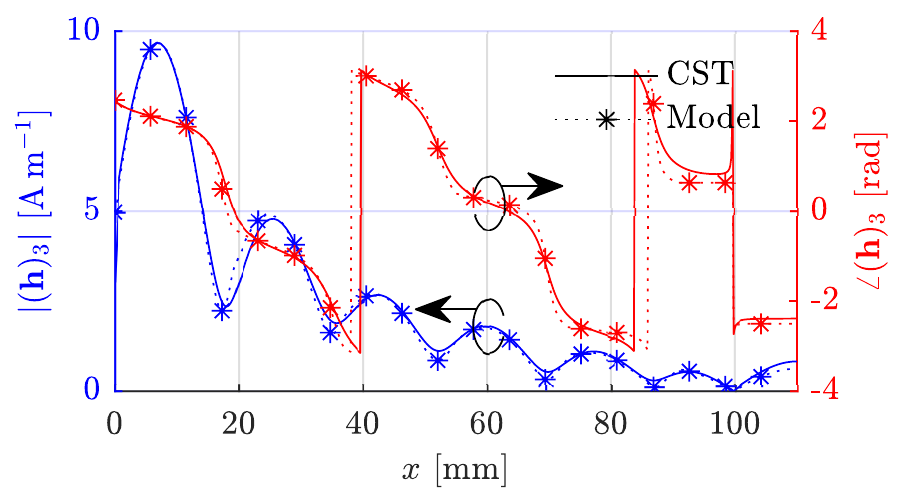}
     \caption{Simulated magnetic field along the center of the waveguide $\left(z=\frac{a}{2} \text{ and }y = \frac{b}{2}\right)$ using both CST and the proposed model.}
     \label{fig:fieldInWaveguide}
 \end{figure}

Fig. \ref{fig:fieldInWaveguide} also displays this behaviour, showing the amplitude and phase of the simulated $z$-component of the magnetic field along the center of the first waveguide ($z=a/2$ and $y = b/2$). The theoretical field is computed by applying the superposition theorem and \eqref{eq:h} with $s_m = (\vec{j}_\text{t})_n$ or  $s_m = (\vec{j}_\text{s})_l$ depending on whether we are computing the field induced by the \ac{RF} chain or the field radiated from a \ac{DMA} element. That is, the field is given by the superposition of the field generated by each of the elements as\footnote{The sum here is only done over the sources and \ac{DMA} elements in contact with the first waveguide}
\begin{align}
    \vec{h}(\vec{r}) = -i\omega\epsilon \bigg(\sum_n^1 G_{e2,zz}^{(w)}\left(\vec{r},\vec{r}_n\right) \left(\vec{j}_\text{t}\right)_n  + \sum_l^{5} G_{e2,zz}^{(w)}\left(\vec{r},\vec{r}_l\right) \left(\vec{j}_\text{s}\right)_l\bigg),
\end{align}
where only the fundamental mode ($n=0$ and $m=1$) of the Green's function $G_{e2,zz}^{(w)}$ is included, i.e., $G_{e2,zz}^{(w)}$ is given as in \eqref{eq:Gw_xx_param}. Also, $\vec{r}_l$ denotes the position of the $l$th \ac{DMA} element, and $\vec{r}_n$ denotes the position of the $n$th \ac{RF} chain input. As observed in Fig. \ref{fig:fieldInWaveguide}, there is a perfect match between the full-wave simulated result and the theoretical expression. 

\begin{figure}[t]
     \centering
     \includegraphics[width=1\linewidth]{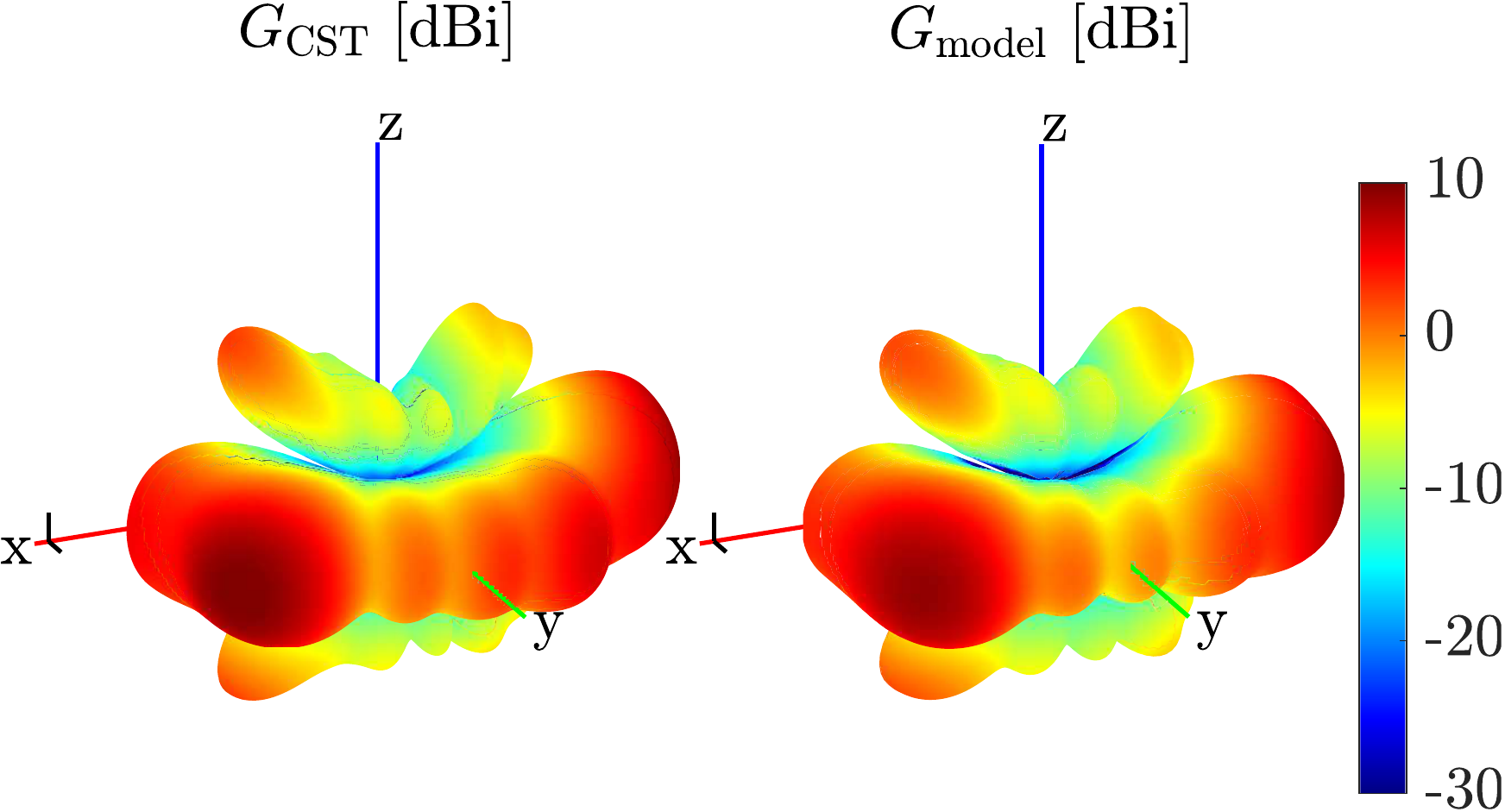}
     \caption{Simulated farfield gain pattern using both CST and the proposed model.} \label{fig:gainPattern}
 \end{figure}

On the other hand, the simulated and theoretical far field gain patterns are illustrated in Fig. \ref{fig:gainPattern}. The simulated pattern is extracted from CST, while the theoretical pattern is derived by calculating again the radiated field in a semi-sphere of radius $R$ encapsulating the \ac{DMA} ($R$ is large enough such that the far field assumption is valid). Specifically, the field $\vec{h}(\vec{r})$ with $\|\vec{r}\|=R$ is calculated by applying superposition and \eqref{eq:h}, where in this case the Green's function is given by $2\mat{G}^{(a)}_{\text{e}2}\left(\vec{r},\vec{r}'\right)$ with $\mat{G}^{(a)}_{\text{e}2}$ as in \eqref{eq:Ga_appendix}\footnote{Note that, in this case we are calculating the radiated field and not the voltage at the receiver $z$ oriented dipole, therefore we need the whole third column of \eqref{eq:Ga_appendix}.} and the currents are those in \eqref{eq:simCurrents}. With $\vec{h}(\vec{r})$, the gain pattern in an arbitrary direction is calculated as \cite{Balanis2016}
\begin{equation}
    G(\vec{r}) = 4\pi R^2\eta\frac{\vec{h}(\vec{r})^H\vec{h}(\vec{r})}{2P_s}, 
\end{equation}
where $P_s = 1$ is the supplied power and $\eta = 120\pi$. 

\begin{figure}[t]
     \centering
     \includegraphics[width=1\linewidth]{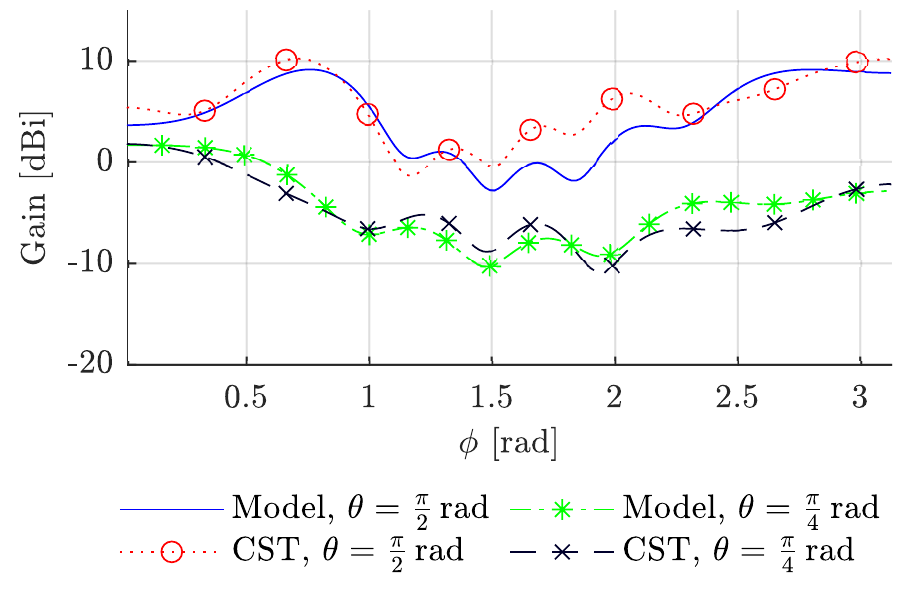}
     \caption{Simulated farfield gain pattern cuts using both CST and the proposed model.} \label{fig:gainPatternCuts}
 \end{figure}

As observed, the theoretical radiation pattern accurately fits the simulated one, capturing the general trend and behaviour of the system, which was precisely the goal of the generic model here proposed. A closer inspection to Fig. \ref{fig:gainPattern} reveals that some minor lobes are not completely captured, being this a consequence of the infinitesimal magnetic dipole model (in the simulated layout we have real antennas) and the fact that we only consider the $z$ component of the magnetic dipole moment. This is better observed in Fig. \ref{fig:gainPatternCuts}, where the $\theta=\frac{\pi}{2}\,\si{\radian}$ and $\theta = \frac{\pi}{4}\,\si{\radian}$ far field cuts are shown. However, it is worth to highlight that we target to capture the overall trend of \ac{DMA} based systems with a low mathematical complexity. 

\section{Conclusions}\label{sec:Conclusions}
This paper has presented a narrowband communication model for \ac{DMA} based systems which accurately captures the physical particularities inherent to these systems. The model, which arises from the underlying electromagnetic theory and hence is consistent with the physics, can be used as a building block and starting point for transmit and receive design optimization and beamforming, allowing to consider important phenomena such that  variable insertion losses due to the backscattering by the \ac{DMA} elements, decreasing wave amplitude due to energy being radiated into free space, and mutual coupling between the radiating \ac{DMA} elements.

The model is rather generic, and its output is an equivalent channel matrix that is easily introduced in the conventional \ac{mMIMO} communication model. All the expressions in the model are given in simple closed-form expressions, and its validity has been verified through full wave simulations, showing a good agreement. Therefore, we can abstract from the underlying electromagnetic theory and work at a signal processing level but keeping consistency with the physics. A \textsc{Matlab} implementation of the proposed model is available at \cite{DMACode2022}.

Examples of future work include the analysis of the degrees of freedom required to limit insertion losses, which arises as a challenge since reconfiguring the \ac{DMA} to obtain a beam in certain direction also changes the power that is supplied to the system. Also, since the amplitude of the field inside the waveguide decreases as the elements radiate part of it out of the cavity, it would be interesting to find the optimum number of elements per area needed to harvest the degrees of freedom available in the channel. Wideband modeling and bandwidth control as well as channel estimation methods are also areas that need to be explored.

\appendices
\label{sec:appendix}
\section{Green's functions}
\subsection{Green's function inside rectangular waveguide}
\label{app:A1}

Inside the finite waveguide, the Green's functions are given as the sum of the Green's function for the infinite waveguide and the scattering from the terminations of the waveguide as \cite[Eq. (4)]{Li1995}
\begin{align}
    \mat{G}_{e2}^{(w)}=\mat{G}_{e2}^{(inf)} + \mat{G}_{e2}^{(sca)}, \label{eq:Ge2W}
\end{align}
where $\mat{G}_{e2}^{(inf)}$ and $\mat{G}_{e2}^{(sca)}$ are given by 
\cite[Eqs. (5) and (6)]{Li1995}. As all dipoles in the model are aligned with the $z$-axis, only the third diagonal component of the Green's function is involved in the mutual admittance calculation, and it is obtained from \cite[Eqs. (5) and (6)]{Li1995} by using \cite[Eqs. (5.19) and (5.20)]{Tai1994} and performing algebraic manipulations, yielding  \eqref{eq:greenWaveguideFull} (placed at the top of next page). In \eqref{eq:greenWaveguideFull}, $\delta_0$ is defined as $\delta_0 = 1$ if $m=1$ or $n=1$ and $\delta_0 = 0$ otherwise.

\begin{figure*}[t]
\begin{align}
G_{e2,zz}^{(w)}=\sum_{n,m}&\frac{-\p{2-\delta_0}\es{\frac{m\pi z}{a}}\ec{\frac{n\pi y}{b}}  \left[\ec{k_x \p{x'+x-S_{\mu}}} + \ec{k_x\p{S_{\mu}-\left|x-x'\right|}}\right]}{ \p{\es{\frac{m\pi z'}{a}} \ec{\frac{n\pi y'}{b}}}^{-1} a b k_x k^2 \p{a^2 n^2 + b^2 m^2}\p{a^2k^2n^2+b^2k_x^2m^2}^{-1}\es{k_x S_\mu}}. \label{eq:greenWaveguideFull}
\end{align}
\hrulefill
\vspace*{4pt}
\end{figure*}

Since the dimensions of the waveguide are chosen such that only the fundamental mode $\text{TE}_{10}$ ($m=1$ and $n=0$) propagates,  $k_x$ is purely real for the $\text{TE}_{10}$ mode  and purely imaginary for the higher order modes ($m>1$ and $n>0$). Therefore, these higher order modes can be neglected as they decrease exponentially with the distance from the source and terminations, and then \eqref{eq:Gw_xx_param} is straightforwardly obtained by setting $m=1$ and $n=0$ in \eqref{eq:greenWaveguideFull}.

On the other hand, when determining the self-admittance of the sources and the self-admittance contribution from the field propagating inside the waveguide for the \ac{DMA} elements, the amplitudes of the higher order modes are not negligible. The higher order modes do however only contribute to the imaginary part which in theory can be adjusted for in the passive termination. As such, for the sake of simplicity, only the fundamental mode ($m = 1$ and $n=0$) is considered in this model.

\subsection{Green's function in free space}
\label{app:A2}
The Green's function in free space $\mat{G}_{\text{e}2}^{(a)}$ can be obtained as the received magnetic field when the medium is excited by an impulsive effective magnetic source. The potential function $\mat{F}\in\mathbb{C}^{3\times 3}$, for an arbitrary source $\mat{S}_m\in\mathbb{C}^{3\times 3}$, is expressed as \cite[Eq. (3.28)]{Balanis2016}
\begin{equation}
    \mat{F} = \frac{\epsilon}{4\pi}\iiint_V \mat{S}_m \frac{\e{-i k \|\vec{r}-\vec{r}'\|_2}}{\|\vec{r}-\vec{r}'\|_2}\dif V'.
\end{equation}
Notice that, in contrast to \cite{Balanis2016}, we are using a matricial notation for both the current source and the potential, similarly as in \cite{Harrington2001}. Therefore, considering $\mat{S}_m = \mat{I}_3\delta(\vec{r}-\vec{r}')${\footnote{\review{Here, $\delta(\vec{r}-\vec{r}')$ is defined such that $\int_0^\infty f(\vec{r})\delta(\vec{r}-\vec{r}')\dif \vec{r} = f(\vec{r}')$ \cite{Tai1994}.}}} leads to 
\begin{equation}
    \mat{F}\left(\vec{r},\vec{r}'\right)=\frac{\epsilon}{4\pi} \mat{I}_3 \frac{\e{-i k \|\vec{r}-\vec{r}'\|_2}}{\|\vec{r}-\vec{r}'\|_2}. \label{eq:Fpotential}
\end{equation}
According to the magnetic dipole model, we neglect the potential generated by the electric source current and hence the radiated magnetic field is only determined by $\mat{F}$ in \cite[Eq. (3.30)]{Balanis2016}. Equating \cite[Eq. (3.30)]{Balanis2016} and \eqref{eq:HIntegral}, and introducing the matrix source distribution, then 
\begin{equation}
     \mat{G}^{(a)}_{\text{e}2}\left(\vec{r},\vec{r}'\right) = \frac{1}{\epsilon} \left(\mat{f}\left(\vec{r},\vec{r}'\right) +\frac{1}{k^2}\nabla \nabla^T \mat{f}\left(\vec{r},\vec{r}'\right)\right). \label{eq:Ga_appendix}
\end{equation}

Finally, introducing \eqref{eq:Fpotential} in \eqref{eq:Ga_appendix} and performing algebraic manipulations yield \eqref{eq:GfreeSpaceFull}.

\section{Free space self-impedance}
\label{app:A3}
Redefining \eqref{eq:GfreeSpaceFull} in terms of the spherical coordinates, i.e., $\Delta z = R\,\ec{\theta}$ (see Fig. \ref{fig:Coordinates}),
yields
\begin{align}
    {G}^{(a)}_{\text{e}2,zz}&\left(\vec{r},\vec{r}'\right) =  \frac{\e{-i k R}}{4 \pi R} \notag \\
    &\times \left( \es{\theta}^2    - i\frac{1-3\ec{\theta}^2}{k R}-\frac{1 - 3\ec{\theta}^2}{k^2 R^2}\right), \label{eq:GreensSpherical}
\end{align}
and the goal is evaluating this expression for $R\to 0$. Whilst the limit in the real part of \eqref{eq:GreensSpherical} diverges, it can be calculated in the imaginary part, which is rewritten as 
\begin{align}
    \Im{{G}^{(a)}_{\text{e}2,zz}} = \frac{f(R)}{g(R)},
\end{align}
where $f\p{R}$ and $g\p{R}$ are
\begin{align}
    f\p{R} = &\p{\p{R^2k^2-3}\ec{\theta}^2 - R^2 k^2 + 1}\es{k R} \notag \\
    &+3 k \p{\ec{\theta}^2 - \frac{1}{3}}R \ec{k R}, \\
    g\p{R} =& 4 \pi R^3 k^2.
\end{align}
Using L'Hôpital's rule, the limit of $\Im{{G}^{(a)}_{\text{e}2,zz}}$ for $R\to 0$ is evaluated as
\begin{align}
    \lim_{R\to 0} \Im{{G}^{(a)}_{\text{e}2,zz}} =\lim_{R\to 0} \frac{f(R)}{g(R)} = \lim_{R\to 0} \frac{\frac{\dif^3f(R)}{\p{\dif R}^3}}{\frac{\dif^3g(R)}{\p{\dif R}^3}} = \frac{-k}{6\pi}.
\end{align}
Introducing this result in \eqref{eq:YssSelf} \review{and taking the limit in \eqref{eq:Gw_xx_param}} readily yield \eqref{eq:Yss_self}.

\section{Solution of integral in equation \eqref{eq:Sigma}} \label{app:solveIntegral}

Instead of calculating \eqref{eq:Sigma} directly, we show that the elements of the covariance matrix can be derived from the mutual admittance in \eqref{eq:YssGeneral}. The mutual admittance between minimum scattering antennas can be derived through direct evaluation of the Green's function as done in sections \ref{sec:DMAanalysis} and \ref{sec:UserCoupling}. Alternatively, it can also be computed through a surface integration of the pointing vector, computing the power escaping out through a surface enclosing the \ac{DMA}. The antenna gain pattern can also be used as an alternative to the pointing vector, yielding the normalized admittance instead. As the waveguide is terminated, no power escapes out through the waveguide and therefore power can only escape out through the upper half-space where $y \geq 0$. 

Taking this into account, for minimum-scattering antennas, the real part of the normalized mutual admittance can be computed through the far field gain pattern as \cite{Wasylkiwskyj1970, Williams2020ICC},
\begin{align}
    \Re{\tfrac{\p{\mat{Y}_\text{ss}}_{n,m}}{\p{\mat{Y}_\text{ss}}_{n,n}}} &= \Int_0^{2\pi} \Int_0^{\pi} \tfrac{G(\theta,\phi)}{4\pi} \e{-i k \nvec{r}^T \p{\vec{r}_n - \vec{r}_m}} \es{\theta}\dif \theta \dif \phi, \label{eq:admittanceintegral}
\end{align}
where $G(\theta,\phi)$ is the gain pattern of the minimum scattering antennas, which for the case of magnetic dipoles on a conductive plane is given by \cite{Balanis2016}
\begin{align}
    G(\theta,\phi) = \begin{cases}
    3\es{\theta}^2 & \text{for } 0 \leq \theta, \phi \leq \pi \\
    0 & \text{otherwise } \\
  \end{cases}. \label{eq:gainPattern}
\end{align}
Inserting \eqref{eq:gainPattern} into \eqref{eq:admittanceintegral} yields
\begin{align}
   \Re{\frac{\p{\mat{Y}_\text{ss}}_{n,m}}{\p{\mat{Y}_\text{ss}}_{n,n}}} \frac{4\pi}{3} &= \Int_0^{\pi} \Int_0^{\pi} \e{-i k \nvec{r}^T \p{\vec{r}_n - \vec{r}_m}} \es{\theta}^3\dif \theta \dif \phi. \label{eq:admittanceintegralfinal}
\end{align}
 
It can be observed that the above integral is exactly the same as in \eqref{eq:Sigma} with $\vec{k}(\theta,\phi) = k\hat{\vec{r}}^T$. Therefore, inserting \eqref{eq:YssGeneral} and \eqref{eq:Yss_self} with ${G}_{\text{e}2,zz} = {G}_{\text{e}2,zz}^{(a)}$ into the left hand side of \eqref{eq:admittanceintegralfinal} and applying the resulting expression in \eqref{eq:Sigma}, yields \eqref{eq:SigmaFinal}.

\bibliographystyle{IEEEtran}
\bibliography{main}

\end{document}